\documentclass[11pt,onecolumn,journal]{IEEEtran}

\pdfoutput=1

\usepackage{setspace}
\usepackage{amsmath,amsopn,amssymb}
\usepackage{graphicx,xspace,color,soul}
\usepackage{longtable,multirow}
\usepackage{array,float}
\usepackage{cite,citesort}

\newtheorem{lemma}{Lemma}

\usepackage{algorithm,algorithmicx}
\usepackage{algpseudocode}
\usepackage{epstopdf}
\usepackage{url}
\usepackage{multirow}
\usepackage[caption=false,font=footnotesize]{subfig}

\usepackage{enumitem}
\usepackage{pgfplots} 
\pgfplotsset{compat=newest}
\pgfplotsset{plot coordinates/math parser=false}
\usetikzlibrary{patterns,decorations.pathreplacing,shapes,arrows,chains}
\usepackage{tikz}

\definecolor{bblue}{HTML}{4F81BD}
\definecolor{rred}{HTML}{C0504D}
\definecolor{ggreen}{HTML}{9BBB59}
\definecolor{ppurple}{HTML}{9F4C7C}
\definecolor{ggray15}{HTML}{D9D9D9}

\newlength\figureheight
\newlength\figurewidth

\begin{document}

\title{Optimal Lagrange Multipliers for Dependent Rate Allocation in Video Coding}

\author{
Ana De Abreu~\IEEEmembership{Student Member,~IEEE},
Gene Cheung~\IEEEmembership{Senior Member,~IEEE},
Pascal Frossard~\IEEEmembership{Senior Member,~IEEE},
Fernando Pereira ~\IEEEmembership{Fellow,~IEEE}
\begin{small}
\thanks{A. De Abreu and P. Frossard are with 
        Signal Processing Laboratory (LTS4), Ecole Polytechnique F\'{e}d\'{e}rale de Lausanne (EPFL), CH-1015 Lausanne, Switzerland
        (e-mail: ana.deabreu@epfl.ch, pascal.frossard@epfl.ch).}
        \thanks{G. Cheung is with 
        National Institute of Informatics, 2-1-2, Hitotsubashi, Chiyoda-ku,
        Tokyo, Japan 101--8430 
        (e-mail: cheung@nii.ac.jp).}
        \thanks{A. De Abreu and F. Pereira are with  
        Instituto Superior T\'{e}cnico - Instituto de Telecomunica\c{c}\~{o}es (IST-IT), 1049-001, Lisbon, Portugal
        (e-mail: fp@lx.it.pt).}

\end{small}
}%
\maketitle
\vspace{0.1in}

\begin{abstract}

In a typical video rate allocation problem, the objective is to optimally distribute a source rate budget among a set of (in)dependently coded data units to minimize the total distortion of all units. 
Conventional Lagrangian approaches convert the lone rate constraint to a linear rate penalty scaled by a multiplier in the objective, resulting in a simpler unconstrained formulation. 
However, the search for the ``optimal" multiplier---one that results in a distortion-minimizing solution among all Lagrangian solutions that satisfy the original rate constraint---remains an elusive open problem in the general setting. 

To address this problem, we propose a computation-efficient search strategy to identify this optimal multiplier numerically. Specifically, we first formulate a general rate allocation problem where each data unit can be dependently coded at different quantization parameters (QP) using a previous unit as predictor, or left uncoded at the encoder and subsequently interpolated at the decoder using neighboring coded units.
After converting the original rate-constrained problem to the unconstrained Lagrangian counterpart, we design an efficient dynamic programming (DP) algorithm that finds the optimal Lagrangian solution for a fixed multiplier. 
Finally, within the DP framework, we iteratively compute neighboring singular multiplier values, each resulting in multiple simultaneously optimal Lagrangian solutions, to drive the rates of the computed Lagrangian solutions towards the bit budget. 
We terminate when a singular multiplier value results in two Lagrangian solutions with rates below and above the bit budget.
In extensive monoview and multiview video coding experiments, we show that, for fixed target rate constraints, our DP algorithm and selection of optimal multipliers on average outperform comparable rate control solutions used in video compression standards such as HEVC that do not skip frames in Y-PSNR.




\end{abstract}


\IEEEpeerreviewmaketitle

\section{Introduction}
\label{sec:intro}

In video coding, rate allocation is the problem of distributing a source bit budget $B$ to a set of (in)dependently coded data units, $v \in \mathcal{V}^o$, in order to minimize the distortion of all units. 
For example, a data unit $v$ can be a video frame predictively coded at a quantization parameter (QP) $q_v$, using a previous frame as a predictor. 
Coding at a larger (coarser) QP requires fewer bits in general but results in a higher quantization distortion. 
In some cases, leaving a unit uncoded at the encoder may be a better rate-distortion (RD) decision; the unit is subsequently interpolated at the decoder using neighboring coded units via techniques such as motion compensated interpolation (MCI) \cite{tekalp95} in monoview video or depth-image based rendering (DIBR) \cite{wang11,kauff07} in multiview video when color and depth maps are available. 
In these cases, the more general rate allocation problem is to first select data units $\mathcal{V} \subseteq \mathcal{V}^o$ for coding, and then select QPs $\mathbf{q} = [q_1, \ldots, q_{|\mathcal{V}|}]$ at which to code the selected units $\mathcal{V}$ to minimize the total distortion $D(\mathcal{V}, \mathbf{q})$ subject to a rate constraint $R(\mathcal{V}, \mathbf{q})$:
\begin{equation}
\min_{\mathcal{V} \subseteq \mathcal{V}^o, \mathbf{q}}
D(\mathcal{V}, \mathbf{q}) 
~~~ \mbox{s.t.} ~ R(\mathcal{V}, \mathbf{q}) \leq B
\label{eq:primal}
\end{equation} 

To address different variants of the rate allocation problem, Lagrangian approaches---where the lone rate constraint is first converted to a linear rate penalty in the objective scaled by a multiplier $\lambda$---are common in the literature \cite{shoham88,kannan94,liu05,kim05,cheung11tip2}. 
This results in a simpler unconstrained problem:
\begin{equation} 
(\mathcal{V}_{\lambda}, \mathbf{q}_{\lambda}) = \arg
\min_{\mathcal{V} \subseteq \mathcal{V}^o, \mathbf{q}} 
D(\mathcal{V}, \mathbf{q}) + \lambda ~  R(\mathcal{V}, \mathbf{q})
\label{eq:dual}
\end{equation}
which in general is easier to solve for a fixed multiplier $\lambda$ \cite{shoham88,kannan94,liu05,kim05,cheung11tip2}.
However, the Lagrangian relaxed problem (\ref{eq:dual}) is inherently not the same as the original rate-constrained problem (\ref{eq:primal}); the difference in distortion between their respective optimal solutions is called a \textit{duality gap} (see Appendix B and \cite{shoham88}). 
To minimize this gap, it is imperative to find the ``optimal" multiplier $\lambda^*$---one that results in a distortion-minimizing solution $(\mathcal{V}_{\lambda^*}, \mathbf{q}_{\lambda^*})$ among all Lagrangian solutions $(\mathcal{V}_{\lambda}, \mathbf{q}_{\lambda})$ to (\ref{eq:dual}) for different $\lambda$ that satisfy $R(\mathcal{V}_{\lambda}, \mathbf{q}_{\lambda}) \leq B$.  
However, given empirical discrete rate and distortion functions $R(\mathcal{V}, \mathbf{q})$ and $D(\mathcal{V}, \mathbf{q})$, the search for this optimal multiplier numerically without resorting to continuous rate and distortion models \cite{wang06,kaaniche14} remains an open problem in the general setting.



To address this problem, in this paper we propose a computation-efficient search strategy to find this optimal multiplier numerically. 
Specifically, we first formulate a general rate allocation problem, where each data unit can be dependently coded at different QPs using a previous coded unit as predictor, or left uncoded at the encoder for interpolation at the decoder using neighboring coded units as reference. 
After converting the original rate-constrained problem to the unconstrained Lagrangian counterpart (\ref{eq:dual}), we design an efficient dynamic programming (DP) algorithm that finds the optimal solution to (\ref{eq:dual}) for a fixed $\lambda$. 

To find the optimal multiplier, we iteratively compute neighboring \textit{singular multiplier values} \cite{shoham88} from the computed DP solution; each singular value $\lambda$ results in multiple simultaneously optimal solutions with different rates, \textit{i.e.} $R(\mathcal{V}_{\lambda}^{(1)}, \mathbf{q}_{\lambda}^{(1)}) \neq R(\mathcal{V}_{\lambda}^{(2)}, \mathbf{q}_{\lambda}^{(2)})$. 
We show that singular values alone lead to all Lagrangian solutions to (\ref{eq:dual}). 
When we obtain solutions $(\mathcal{V}_{\lambda^*}^{(1)}, \mathbf{q}_{\lambda^*}^{(1)})$ and $(\mathcal{V}_{\lambda^*}^{(2)}, \mathbf{q}_{\lambda^*}^{(2)})$ corresponding to singular value $\lambda^*$ with rates $R(\mathcal{V}_{\lambda^*}^{(1)}, \mathbf{q}_{\lambda^*}^{(1)}) \leq B \leq R(\mathcal{V}_{\lambda^*}^{(2)}, \mathbf{q}_{\lambda^*}^{(2)})$, we prove that $(\mathcal{V}_{\lambda^*}^{(1)}, \mathbf{q}_{\lambda^*}^{(1)})$ is the distortion-minimizing Lagrangian solution, and declare $\lambda^*$ as the optimal multiplier.
To the best of our knowledge, no previously proposed Lagrangian multiplier searches \cite{shoham88,kannan94,liu05,kim05,cheung11tip2} provide this theoretical claim in our general setting\footnote{\cite{shoham88} was a seminal bit allocation work and proposed the first singular value search strategy but only for independently coded data units.}. 



Experimental results illustrate good performance of our proposed rate allocation algorithm with optimal Lagrange multiplier selection when data units are independently or predictively coded in both monoview and multiview video sequence compression. Specifically, we show that our bit allocation strategy on average outperforms comparable rate control solutions adopted in the reference softwares of current monoview and multiview video standards, namely HEVC  \cite{bin12} and 3D-HEVC \cite{woong13}, that do not skip frames in Y-PSNR. 

The paper is organized as follows. 
We first review related work in Section \ref{sec:related} and formulate our dependent rate allocation problem in Section \ref{sec:formulate}. 
We then describe a DP algorithm that solves the rate-constrained problem optimally but in exponential time in Section \ref{sec:lagrange}. 
To reduce complexity, we convert the problem to the Lagrangian relaxed version and propose a corresponding polynomial-time DP algorithm for a fixed multiplier. 
In Section \ref{sec:singular}, we discuss an efficient search methodology to identify the optimal multiplier based on the computation of neighboring singular multiplier values. 
Finally, we present experimental results and conclusion in Section \ref{sec:results} and \ref{sec:conclude}, respectively.

\section{Related Work}
\label{sec:related}


\subsection{Continuous RD Function Modeling}


A simple approach to obtain rate and distortion functions is to apply different QPs to encode each data unit and compute the resulting rate and distortion values. 
Then, these empirical RD points are fitted into mathematical functions for each particular sequence to derive an RD model \cite{siwei12,stuhlmuller00}. 
However, in these type of models some parameters have to be estimated for each given video sequence and therefore they cannot be easily generalized. 

Alternatively, it is possible to theoretically derive the different parameters of the RD model under simplifying assumptions. Several RD models using well understood exponential functions have been proposed \cite{wang06,kaaniche14}.
\cite{wang06} considered both a Laplacian and a Generalized Gaussian (GG)  distribution in their RD model for a wavelet video coding. 
\cite{kaaniche14} adopted the Bernoulli Generalized Gaussian (BGG) model for both rate and distortion functions. 
In general, these approaches suffer from: i) modeling errors due to idealized model inaccuracy; and ii) continuous approximation error since the problem (selecting QPs from a finite set) is inherently discrete.
In contrast, we take an empirical approach and solve the inherent discrete problem of selecting data units and QPs for coding directly, and thus do not suffer from modeling errors. 


\subsection{Constrained Formulation via Dynamic Programming}


Addressing directly the discrete rate-constrained bit allocation problem, one common approach is DP.
For example, assuming independently coded data units, \cite{ortega98} constructed a tree to represent all possible solutions: a node at a stage $i$ of the tree represents a particular selection of QPs $\{q_1, \ldots, q_i\}$ for data unit $1$ to $i$. 
If two different nodes at the same stage $i$ have the same accumulated rate from unit $1$ to $i$, then the one with the larger accumulated distortion would be pruned. 
As we will show in Section \ref{sec:formulate}, the complexity of this type of DP algorithms is \textit{pseudo-polynomial} or exponential time.
If predictive coding is assumed, complexity is even higher. 


To jointly select predictor frames and QoS levels for encoding and protection of different video frames during network streaming, \cite{cheung07transcsvt} proposed an integer rounding approach to reduce complexity of a DP algorithm, where DP tables used to store computed local solutions were scaled down to reduce the number of table entries. The authors derived a performance bound for the proposed reduced-complexity DP algorithm; however, this bound gets progressively worse as the scale factor increases. 

For multiview, assuming independently coded units, \cite{anavcip2014} considered a uniform rate allocation among views in a multiview video system, and proposed a DP-based algorithm to select the views for encoding and transmission such that the expected distortion among encoded and synthesized views is minimized given a rate budget. This work is extended in \cite{ana2015} where both the views and the coding rates for each selected view are selected to minimize distortion in a rate-constrained scenario. Due to high complexity, a greedy DP algorithm was proposed. 
There is no performance guarantee for the greedy DP algorithm, however, and thus the obtained solution can be arbitrarily far from the optimal solution.



\subsection{Discrete Lagrangian Formulation}

Instead of the originally posed rate-constrained bit allocation problems, the Lagrangian approach is often used. 
\cite{kannan94} proposed a trellis-based algorithm to find the optimal Lagrangian solution for predictively coded frames in traditional monoview video. 
A suitable Lagrange multiplier was found by sweeping from 0 to $\infty$; it is not clear how this can be done efficiently, and what is the termination condition when a sufficiently good multiplier value is found.
\cite{Li2016} also adopted a Lagrangian approach, where the value of the Lagrange multiplier $\lambda$ is empirically modeled as a function of the rate and the distortion when data units are predictively encoded. 
There is no guarantee that this ad-hoc approach will result in the \textit{optimal} multiplier value upon termination, however.


To allocate bits among independent quantizers, \cite{shoham88} defined the notion of singular multiplier values---multipliers with multiple simultaneously optimal Lagrangian solutions---and proposed to iteratively search through neighboring singular multiplier values until a terminating condition (resulting rates of Lagangian solutions being very close to bit budget) is met. 
Extending \cite{shoham88}, \cite{cheung00} addressed a similar rate allocation problem with two rate constraints, also using the notion of singular values, in order to achieve an optimal distribution of source and channel bits among wavelet subbands for transmission of scalable video over noisy channels. 
Our algorithm adopts the singular multiplier value concept \cite{shoham88} and extends it to the case where data units can be predictively coded or left uncoded entirely for subsequent interpolation at the decoder.

\subsection{Multiview Rate Allocation}

Rate allocation problems have also been investigated recently in multiview video coding applications. 
\cite{kim05} extended the trellis optimization approach in \cite{kannan94} to multiview video with predictive coding. 
The authors considered a system where views can be skipped at the encoder and eventually synthesized using both texture and depth maps at the decoder, and optimized only QPs of the texture maps. 
In a different framework, \cite{cheung11tip2} tackled the bit allocation problem for both texture and depth data such that the distortion of the camera views and a set of synthetic views, reconstructed at the decoder, is minimized. 
The authors optimized both the set of coded views and their QPs and adopted a trellis-based solution for an effective search of the optimal coding solution. Both \cite{kim05} and \cite{cheung11tip2} employed the Lagrangian approach, but no mathematically rigorous strategy was proposed to search for a suitable multiplier value. 
We will show that our proposal can be applied to multiview coding scenarios also.




\section{Rate Allocation Framework}
\label{sec:formulate}
In a classical rate allocation problem, the objective is to minimize the total distortion of a set of data units, each of which may be independently or predictively coded, subject to a rate budget constraint. We first describe the general coding system under consideration. Then, we formulate a rate-constrained bit allocation problem for predictively coded data units as a discrete optimization problem. We present examples that show how our formulation can be applicable in different practical scenarios. Finally, we prove that the formulated problem is NP-hard.

\subsection{System Model}
\label{sec:SM}

We consider a general coding scenario where we seek to allocate a total bit budget $B$ to an ordered set of $V$ data units, $\mathcal{V}^o=\{1, 2, \ldots, V \}$. Examples include consecutive frames in monoview video, or neighboring views in a multiview image sequence. We define $\mathcal{V}= \{v_1, v_2, \cdots, v_N \}$, where $\mathcal{V} \subseteq \mathcal{V}^o$,  as the subset of $N$ units selected for coding, where $N \leq V$. We assume that a unit $v$ can be left uncoded at the encoder, and later interpolated at the decoder using the two surrounding coded units $v_L$ and $v_R$, where $v_L <  v < v_R$ and $v_L, v_R \in \mathcal{V}$. As the boundary units cannot be interpolated at the decoder in the same manner, they are always selected for coding, \textit{i.e.}, $v_1=1$ and $v_N=V$ are always coded.

Each unit $v_n \in \mathcal{V}$ selected is coded using a QP $q_{v_n} \in \mathcal{Q}$, where $\mathcal{Q}$ is a discrete set of possible QPs for a given encoder. Denote by $\mathbf{q}_\mathcal{V}$ the set of chosen QPs for the units in $\mathcal{V}$. Assuming that predictive coding is used to code neighboring units, unit $v_n$ coded with QP $q_{v_n}$ using as predictor unit $v_{n-1}$ coded with QP $q_{v_{n-1}}$ has a rate $r_{v_n}(v_{n-1}, q_{v_{n-1}}, q_{v_n})$ and a distortion $\Delta_{v_n}(v_{n-1}, q_{v_{n-1}}, q_{v_n})$. Note that if $v_{n-1}$ and $v_n$ are not consecutive data units in $\mathcal{V}^o$, then the uncoded intermediate units between coded $v_{n-1}$ and $v_n$ that are interpolated at the decoder must be included in the distortion computation. Hence distortion term $\Delta_{v_n}(v_{n-1}, q_{v_{n-1}}, q_{v_n})$ accounts for the distortion of all interpolated units in the range $(v_{n-1}, v_n)$ \textit{and} the distortion of the coded unit $v_n$. Mathematically, $\Delta_{v_n}(v_{n-1},q_{v_{n-1}}, q_{v_n})$ can be written as:

\begin{equation}\label{eq:Delta_VR}
\Delta_{v_n}(v_{n-1},q_{v_{n-1}}, q_{v_n})  =
\sum_{\substack{v \in \mathcal{V}^o  \\ v_{n-1} < v \leq v_n \\ q_{v_{n-1}}, q_{v_n} \in \mathcal{Q}}}  d_v(v_{n-1}, v_n, q_{v_{n-1}}, q_{v_n})
\end{equation}
where $d_v(.)$ is the distortion of unit $v$, $v \in \mathcal{V}^o$, interpolated using reference units $v_{n-1}$ and $v_n$ coded using QPs $q_{v_{n-1}}$ and $q_{v_n}$, respectively. If $v=v_n$, then $d_v(.)$ corresponds to the distortion of coded unit $v_n$ using $v_{n-1}$ for prediction. 

The first unit $v_1$ in $\mathcal{V}$ is independently encoded and its distortion depends only on its own QP. For this particular case, Eq. (\ref{eq:Delta_VR}) can be re-written as:


\begin{equation}\label{eq:Delta_V1R1}
\Delta_{v_1}(q_{v_1}) =  d_{v_1}(q_{v_1})
\end{equation}

More general definitions for predictive coding is also possible \cite{kannan94}, where the rate and distortion functions depend on the QPs of \textit{all} previous coded data units. However, for complexity reasons, we assume that rate $r_{v_n}$ and distortion $\Delta_{v_n}$ depend only on QP $q_{v_{n-1}}$ of previous unit $v_{n-1}$ used for prediction. This is a good approximation in practical predictive coding, as shown in \cite{cheung11tip2}.




\subsection{Problem Formulation}\label{sec:pf} 

With the above definitions, our objective is to find the optimal subset of data units $\mathcal{V}^* = \{v_1, v_2, \cdots, v_N \} \subseteq \mathcal{V}^o$ along with their corresponding QPs $\mathbf{q}_\mathcal{V}^*=[q_{v_1}, q_{v_2}, \cdots, q_{v_N}]$ such that the aggregate distortion at the decoder is minimized, subject to a bit budget constraint $B$. The optimization problem can be defined as follows:
\begin{equation}
\label{eq:Problem}
(\mathcal{V}^*, \mathbf{q}_\mathcal{V}^*)   = 
\arg \min_{\substack{\mathcal{V} \subseteq \mathcal{V}^o \\ q_{v_n} \in \mathcal{Q}}}~~\,
 \Delta_{v_1}(q_{v_1}) +  \sum_{n=2}^{|\mathcal{V}|} \Delta_{v_n}(v_{n-1},q_{v_{n-1}}, q_{v_n}) \nonumber
 \end{equation}
 
\begin{equation}
\mbox{s.t.} ~~ r_{v_1}(q_{v_1}) + \sum_{n=2}^{|\mathcal{V}|} r_{v_n}(v_{n-1}, q_{v_{n-1}}, q_{v_n}) \leq B 
\end{equation}
where $\Delta_{v_1}(q_{v_1})$ and $r_{v_1}(q_{v_1})$ are the distortion and rate for the first selected unit $v_1$, and $\Delta_{v_n}(.)$ and $r_{v_n}(.)$ are the distortion and rate for a predictively coded unit $v_n$, as described above.




\subsection{Applications}

Our formulation in (\ref{eq:Problem}) is sufficiently general for application to different monoview and multiview video rate allocation scenarios. 
We list a few illustrative examples below.

\begin{itemize}

\item \emph{Scenario I: QP selection for independent coded images}. If frames in a monoview video or views in a multiview image sequence are independently coded for maximum random access \cite{shoham88}, then our formulation (\ref{eq:Problem}) is applicable to optimal selection of QPs, where the rate and distortion of each data unit (image) do not depend on the previous one, \textit{i.e.}, $r_{v_n}(v_{n-1},q_{v_{n-1}},q_{v_n})= r_{v_n}(q_{v_n})$ and $\Delta_{v_n}(v_{n-1},q_{v_{n-1}},q_{v_n})= d_{v_n}(q_{v_n})$.

\item \emph{Scenario II: QP selection for differentially coded images}. If each frame in monoview video or view in a multiview image sequence is differentially coded using a previous predictor frame, then (\ref{eq:Problem}) can be used, where all units in $\mathcal{V}^o$ are chosen for coding, for optimal selection of QPs \cite{kannan94,cheung11tip2}. Note that (\ref{eq:Problem}) is not applicable to bi-directional prediction like B-frames.

\item \emph{Scenario III: Selection of images for coding}. If a consistent quality requirement dictates that all images should be coded at the same pre-defined QP \cite{anavcip2014}, (\ref{eq:Problem}) can be used to select a subset of images in monoview video or views in a multiview image sequence to minimize aggregate distortion that includes interpolated images at the decoder.

\item \emph{Scenario IV: Mode selection in a Group of Blocks}. Instead of images, data units can represent code blocks in an image. Given a fixed QP, (\ref{eq:Problem}) can be used to select the optimal coding modes for a group of block (GoB) for a given rate constraint, where $\mathcal{Q}$ now represents possible modes a block can take on \cite{sullivan98,cheung01}. 

\end{itemize}





\subsection{NP-Hardness Proof} \label{sec:NP_hard}  

We prove that our formulated bit allocation problem (\ref{eq:Problem}) is NP-hard via reduction from a well-known NP-hard problem---\emph{Knapsack} (KS) \cite{garey99}. The binary decision version of KS, which is NP-complete, can be described as follows: 

\vspace{0.05in}
\noindent 
\emph{Binary Decision Problem of KS} -- Given a set of $M$ items, each with non-negative weight $w_m$ and profit $c_m$, and a knapsack of capacity $W$, does there exist a subset of items with total weight $\leq W$, such that the total profit is at least $\bar{C}$?

\vspace{0.05in}
To prove NP-hardness of (\ref{eq:Problem}), we consider a more specific problem where each unit $v$, if chosen for coding, can only be independently coded at QP $q$.
We reduce the binary decision version of this simplified problem from the KS decision problem as follows. 
First, we construct two boundary units $v_0$ and $v_{M+1}$ and $M$ intermediate data units corresponding to $M$ items in KS. The distortion of not coding any intermediate unit is $D$ no matter what surrounding units are used as reference for interpolation. 
Coding the two boundary units at QP $q$ results in rate $2$ and distortion $0$.
Coding an intermediate unit $v$ at QP $q$ results in rate $w_v$ and distortion $D - c_v$. 
The binary decision problem is: does there exist a subset $\mathcal{V}$ of  units selected for coding (each at QP $q$) such that the distortion is no larger than $M D - \bar{C}$, given a rate budget $W + 2$? If the answer is yes, then the chosen subset $\mathcal{V}$ of units in the solution, excluding the two boundary units, has a corresponding subset of items in KS with total weight no larger than $W$ and total profit at least $\bar{C}$. Hence the problem is no easier than the KS decision problem, and thus is also NP-complete. Therefore the optimization version of the problem is NP-hard. Since the specific problem is already NP-hard, the more general problem (\ref{eq:Problem}) is no easier, and hence is also NP-hard.

\section{DP Algorithm for Lagrangian Problem}
\label{sec:lagrange}
We first present an algorithm based on DP that returns an optimal solution to the constrained problem in (\ref{eq:Problem}). We then show that the algorithm complexity is exponential. Next, we present an alternative DP algorithm that solves the corresponding Lagrangian relaxed problem in polynomial time for a fixed Lagrangian multiplier.




\subsection{Constrained DP Algorithm}
\label{subsec:DP}

To solve the problem in (\ref{eq:Problem}) optimally, we derive a DP algorithm that recursively divides the original problem into smaller sub-problems. When a sub-problem is solved, its solution is stored inside an entry in a DP table, so that subsequent calls to the same sub-problem can simply look up the solved solution in the table \cite{Cormen2009}. 

Denote by $\Phi_{v_{n}}(q_{v_{n}}, \bar{B})$ the minimum distortion sum for data units from $v_{n}+1$ to $V$, given that $v_{n}$ is coded with QP $q_{v_{n}}$, and there is an available bit budget of $\bar{B}$, $\bar{B} \leq B$, to code the remaining units $v_{n+1}, \ldots, v_{N}$. This distortion sum $\Phi_{v_{n}}(q_{v_{n}}, \bar{B})$ can be recursively written as:

\vspace{-0.1in}
\begin{small}
\begin{align}\label{eq:DP}
\Phi_{v_{n}}(q_{v_{n}}, & \bar{B}) =  \underset{\stackrel{v_{n+1} \in \mathcal{V}^o \;|\; v_{n+1} > v_n}{q_{v_{n+1}} \in \mathcal{Q}}} {\mathrm{min}} ~  \Delta_{v_{n+1}}(v_n,q_{v_n},q_{v_{n+1}}) \nonumber \\ 
& + \, \mathbf{1}(v_{n+1} < V)\;
\Phi_{v_{n+1}}(q_{v_{n+1}}, \bar{B}-r_{v_{n+1}}(v_n, q_{v_n}, q_{v_{n+1}}))
\end{align}
\end{small}\noindent
where $\mathbf{1}(c)$ is an indicator function that returns $1$ if the clause $c$ is true, and $0$ otherwise.

In words, (\ref{eq:DP}) selects the next unit $v_{n+1}$ to code at QP $q_{v_{n+1}}$, resulting in distortion $\Delta_{v_{n+1}}(v_n, q_{v_n}, q_{v_{n+1}})$ for data units from $v_n+1$ to $v_{n+1}$ inclusively. This selection means that the bit budget is reduced to $\bar{B} - r_{v_{n+1}}(v_n, q_{v_n}, q_{v_{n+1}})$ for coding of the remaining units. If $v_{n+1} = V$, meaning it is the right boundary unit in $\mathcal{V}^o$, then the recursive term in (\ref{eq:DP}) is not necessary. 


Since the first unit $v_1 = 1$ is always selected for coding, the computation in (\ref{eq:DP}) can be solved via the following initial call:
\begin{equation}
\label{eq:DP_v1}
\underset{\substack{q_{1} \in \mathcal{Q}}} {\mathrm{min}} ~  \Delta_{1}(q_{1}) +  \Phi_{1}(q_{1},  B - r_{1}(q_1)),
\end{equation}

The complexity of the DP algorithm is bounded by the size of the DP table multiplied by the complexity of computing each entry: $O(V^2 Q^2 B)$. This is polynomial in $B$, but $B$ is encoded in $\log_2 (B)$ bits as input to the algorithm, and thus the algorithm is exponential in the size of the input. The complexity is also called \textit{pseudo-polynomial time} in the complexity literature \cite{papadimitriou98}.




\subsection{Lagrangian DP Algorithm}
\label{subsec:L_DP}

To reduce the complexity of the constrained DP algorithm in (\ref{eq:DP}), we seek to eliminate the rate dimension $B$ in the DP table. Towards that goal, we consider a Lagrangian relaxation of our constrained problem in (\ref{eq:Problem}), where we move the rate consideration from the constraint to the objective function, resulting in a rate-distortion (RD) formulation:
\begin{align}
(\mathcal{V}^*, \mathbf{q}_\mathcal{V}^*) = & 
\arg \min_{\substack{\mathcal{V} \subseteq \mathcal{V}^o \\ q_{v_n} \in \mathcal{Q}}}~~\,
\Delta_{v_1}(q_{v_1}) +
\sum_{n=2}^{|\mathcal{V}|} \Delta_{v_n}(v_{n-1},q_{v_{n-1}}, q_{v_n}) 
\nonumber \\
 & + \lambda \left(
r_{v_1}(q_{v_1}) +
\sum_{n=2}^{|\mathcal{V}|} r_{v_n}(v_{n-1}, q_{v_{n-1}}, q_{v_n}) \right) 
\label{eq:DualProblem}
\end{align}
where the multiplier $\lambda > 0$ is a parameter that weighs the importance of rate against distortion. 

To solve (\ref{eq:DualProblem}) for a given $\lambda$, we follow a similar procedure. We first denote $\Phi_{v_n}(q_{v_n})$ as the minimum RD cost for data units from $v_n+1$ to $V$ inclusively, given that $v_n$ is coded with QP $q_{v_n}$. $\Phi_{v_n}(q_{v_n})$ can be defined recursively as: 
\begin{align}
\label{eq:L_DP}
 & \Phi_{v_n}(q_{v_n}) = \underset{\substack{\stackrel{v_{n+1} \in \mathcal{V}^o \;|\; v_{n+1} > v_n}{q_{v_{n+1}} \in \mathcal{Q}}}} {\mathrm{min}} ~ \Delta_{v_{n+1}}(v_n,q_{v_n},q_{v_{n+1}}) \nonumber \\
& + \lambda ~ r_{v_{n+1}}(v_n, q_{v_n}, q_{v_{n+1}}) +  \mathbf{1}(v_{n+1} < V)\; \Phi_{v_{n+1}}(q_{v_{n+1}})
\end{align}



Similar analysis as done for the contrained DP algorithm can show that, for a given $\lambda$, the algorithm (\ref{eq:L_DP}) has complexity $O(V^2 Q^2)$, which is polynomial time.

We now discuss the relationship between the constrained problem in (\ref{eq:Problem}), solvable via (\ref{eq:DP}), and its Lagrangian relaxed version in (\ref{eq:DualProblem}), solvable via (\ref{eq:L_DP}). 
Denote by $(\mathcal{V}_\lambda, \mathbf{q}_\lambda)$ an optimal solution of (\ref{eq:DualProblem}) for a given $\lambda$, with resulting distortion and rate $D(\mathcal{V}_\lambda, \mathbf{q}_\lambda)$ and $R(\mathcal{V}_\lambda, \mathbf{q}_\lambda)$, respectively. 
One can show that, if there exists a multiplier $\lambda^*$ such that $R(\mathcal{V}_{\lambda^*}, \mathbf{q}_{\lambda^*}) = B$, then solution $(\mathcal{V}_{\lambda^*}, \mathbf{q}_{\lambda^*})$ is also an optimal solution of (\ref{eq:Problem}). 
The proof is given in Appendix \ref{Ap:Proof_of_Optimality} for the sake of completeness. 

Because $R(\mathcal{V}_\lambda, \mathbf{q}_\lambda)$ is discrete, there may not exist a multiplier $\lambda$ such that $R(\mathcal{V}_\lambda, \mathbf{q}_\lambda) = B$.  In this case, we can pick a value $\lambda = \lambda_1$  with a corresponding Lagrangian solution $(\mathcal{V}_{\lambda_1}, \mathbf{q}_{\lambda_1})$, $R(\mathcal{V}_{\lambda_1}, \mathbf{q}_{\lambda_1}) < B$, as an approximate solution to (\ref{eq:Problem}) with the following performance bound.
Given two solutions of (\ref{eq:DualProblem}) $(\mathcal{V}_{\lambda_1}, \mathbf{q}_{\lambda_1})$ and $(\mathcal{V}_{\lambda_2}, \mathbf{q}_{\lambda_2})$, using respective multipliers $\lambda_1$ and $\lambda_2$, with resulting rates $R(\mathcal{V}_{\lambda_1}, \mathbf{q}_{\lambda_1}) < B < R(\mathcal{V}_{\lambda_2}, \mathbf{q}_{\lambda_2})$, the difference in distortion between Lagrangian solution $(\mathcal{V}_{\lambda_1}, \mathbf{q}_{\lambda_1})$ and the \textit{true} optimal solution $(\mathcal{V}^o, \mathbf{q}^o)$ of (\ref{eq:Problem}) is bounded as:
\begin{equation}
| D(\mathcal{V}_{\lambda_1}, \mathbf{q}_{\lambda_1}) - D(\mathcal{V}^o, \mathbf{q}^o) | \leq 
| D(\mathcal{V}_{\lambda_1}, \mathbf{q}_{\lambda_1}) - D(\mathcal{V}_{\lambda_2}, \mathbf{q}_{\lambda_2}) |
\label{eq:bound}
\end{equation}

The proof is given in Appendix \ref{Ap:Performance_Bound}. Clearly, the bound in (\ref{eq:bound}) is tightest when the difference in distortion between the two Lagrangian solutions is the smallest. 
\textit{We propose an efficient algorithm to find Lagrange multipliers such that the resulting Lagrangian optimal solutions yield the tightest bound possible with respect to the original constrained problem.}

\section{Search for the Optimal Lagrange Multiplier}
\label{sec:singular}
We propose a methodology to identify the ``optimal" Lagrange multiplier value via an iterative search. By ``optimal", we mean a multiplier value that yields a pair of Lagrangian optimal solutions to (\ref{eq:DualProblem}) with the tightest distortion bound (\ref{eq:bound}) possible with respect to the true optimal solution in (\ref{eq:Problem}). We first review the notion of \textit{singular value} of Lagrange multiplier, introduced in the context of rate allocation problems in \cite{shoham88}. We then discuss our methodology in the following two subsections (neighboring singular value computation and initialization).

\subsection{Singular Values of Lagrange Multiplier}

\begin{figure}[b]
\centering
\includegraphics[width=2.8in]{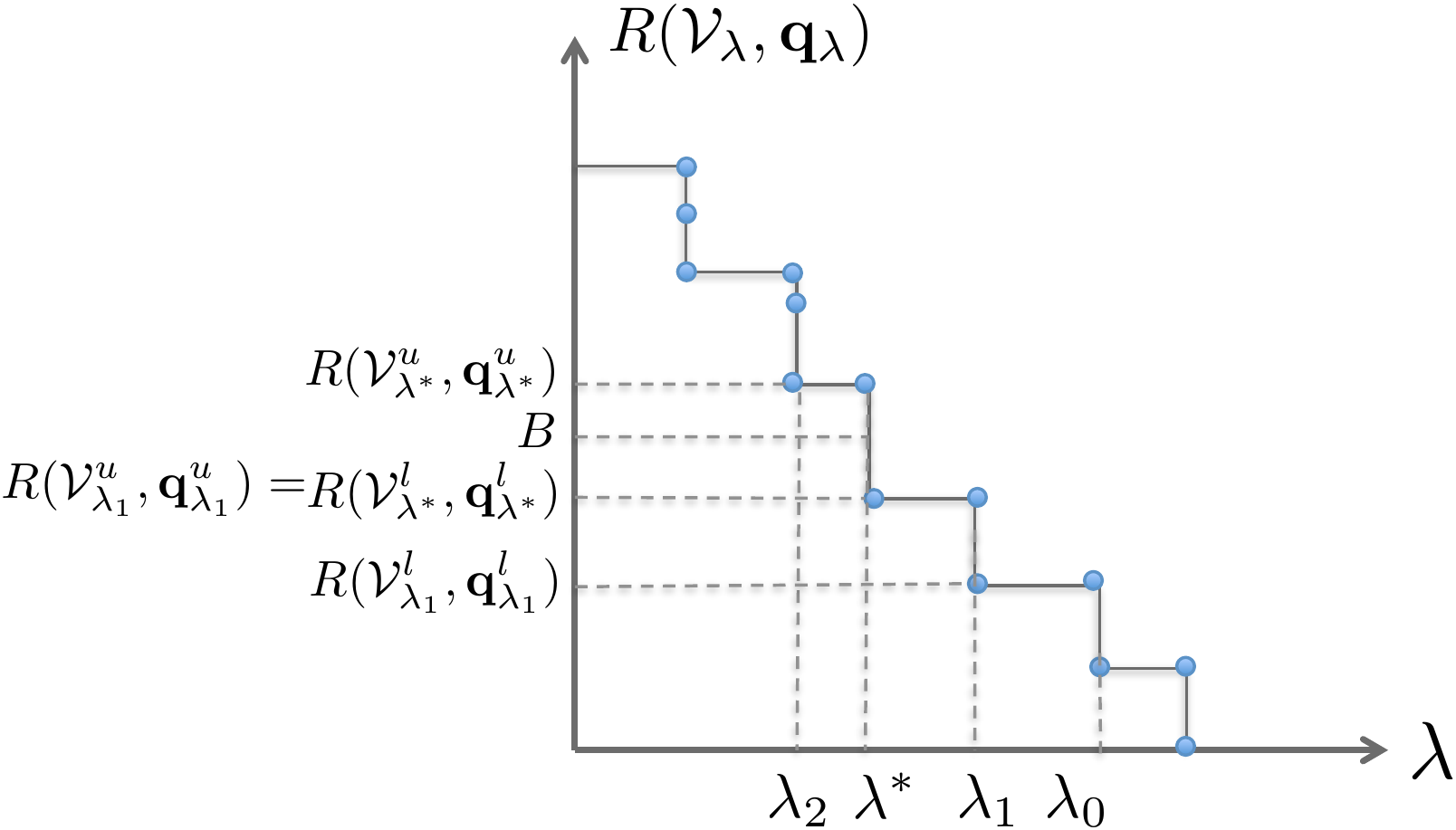}
\vspace{-0.1in}
\caption{Rate $R(\mathcal{V}_{\lambda}, \mathbf{q}_{\lambda})$ of optimal solution $(\mathcal{V}_{\lambda}, \mathbf{q}_{\lambda})$ to (\ref{eq:DualProblem}) as function of multiplier $\lambda$. Singular values are $\lambda_2, \lambda^*, \lambda_1, \lambda_0$, etc. The singular value $\lambda^*$ produces a pair of Lagrangian solutions $(\mathcal{V}^l_{\lambda^*}, \mathbf{q}^l_{\lambda^*})$ and $(\mathcal{V}^u_{\lambda^*}, \mathbf{q}^u_{\lambda^*})$ with the tightest distortion bound (\ref{eq:bound}) with respect to the optimal solution in (\ref{eq:Problem}) with rate constraint $B$.} 
\label{fig:R_lambda}
\end{figure}

We first observe that because rate $R(\mathcal{V}_\lambda, \mathbf{q}_\lambda)$ is discrete, there are distinct $\lambda$ values at which the optimal solutions to (\ref{eq:DualProblem}) are not unique; these are called \textit{singular values} of Lagrange multiplier \cite{shoham88}. As an example, in Fig.~\ref{fig:R_lambda} where rate $R(\mathcal{V}_{\lambda}, \mathbf{q}_{\lambda})$ of optimal solutions $(\mathcal{V}_{\lambda}, \mathbf{q}_{\lambda})$ to (\ref{eq:DualProblem}) is plotted against multiplier $\lambda$, we see that at singular value $\lambda_1$, there are two simultaneous optimal solutions to (\ref{eq:DualProblem}), resulting in two different rates $R(\mathcal{V}^l_{\lambda_1}, \mathbf{q}^l_{\lambda_1})$ and $R(\mathcal{V}^u_{\lambda_1}, \mathbf{q}^u_{\lambda_1})$.
Singular multiplier values have two important properties:

\begin{enumerate}
\item Two neighboring singular values share one common optimal solution to (\ref{eq:DualProblem}).
\item Multipliers $\lambda$ between two neighboring singular values produce the same optimal solution as the shared solution of the two singular values. 
\end{enumerate}
These two properties are discussed extensively in \cite{shoham88}. As an example, in Fig. \ref{fig:R_lambda} neighboring singular values $\lambda^*$ and $\lambda_1$ share an optimal solution $(\mathcal{V}^l_{\lambda^*}, \mathbf{q}^l_{\lambda^*})$ to (\ref{eq:DualProblem}), and multipliers $\lambda$ between these two singular values will produce the same optimal solution.
These two properties imply the following important corollary:  singular values alone produce \textit{all} solutions to (\ref{eq:DualProblem}) as $\lambda$ varies from 0 to $\infty$. Thus, it is sufficient to examine only Lagrangian solutions of singular values in order to find the best multiplier value. 

Moreover, it is known \cite{shoham88} that rate $R(\mathcal{V}_\lambda, \mathbf{q}_\lambda)$ is monotonically non-increasing with respect to $\lambda$ as shown in Fig. \ref{fig:R_lambda}. 
Suppose now that a singular value $\lambda^*$ has corresponding optimal Lagrangian solutions $(\mathcal{V}_{\lambda^*}^l, \mathbf{q}_{\lambda^*}^l)$ and $(\mathcal{V}_{\lambda^*}^u, \mathbf{q}_{\lambda^*}^u)$ where
$R(\mathcal{V}_{\lambda^*}^l, \mathbf{q}_{\lambda^*}^l) \leq B \leq R(\mathcal{V}_{\lambda^*}^u, \mathbf{q}_{\lambda^*}^u)$. That means that any other singular value will yield Lagrangian solutions with rates either smaller or larger than this pair of solutions due to monotonicity of rate $R(\mathcal{V}_\lambda, \mathbf{q}_\lambda)$. Thus this pair of solutions $(\mathcal{V}_{\lambda^*}^l, \mathbf{q}_{\lambda^*}^l)$ and $(\mathcal{V}_{\lambda^*}^u, \mathbf{q}_{\lambda^*}^u)$ are the Lagrangian solutions that produce the tightest distortion bound (\ref{eq:bound}) possible, and $\lambda^*$ is the optimal Lagrange multiplier value.

Procedurally, monotonicity also means that in an iterative search for the optimal singular value $\lambda^*$, one only needs to increase / decrease the current $\lambda$  by examining the rate of the corresponding optimal solution: decrease $\lambda$ if $R(\mathcal{V}_\lambda, \mathbf{q}_\lambda) < B$ and increase it otherwise. 



\subsection{Procedure to compute Neighboring Singular Values}

One strategy to search for the optimal singular value is to march through neighboring singular values in the direction of bit budget $B$ until the pair of optimal solutions to (\ref{eq:DualProblem}) corresponding to singular value $\lambda^*$, $(\mathcal{V}_{\lambda^*}^l, \mathbf{q}_{\lambda^*}^l)$ and $(\mathcal{V}_{\lambda^*}^u, \mathbf{q}_{\lambda^*}^u)$, have rates satisfying $R(\mathcal{V}_{\lambda^*}^l, \mathbf{q}_{\lambda^*}^l) \leq B \leq R(\mathcal{V}_{\lambda^*}^u, \mathbf{q}_{\lambda^*}^u)$. For example, in Fig.\;\ref{fig:R_lambda}, after testing $\lambda_0$ then $\lambda_1$ successively, we arrive at the optimal singular value $\lambda^*$. 
Thus the challenge is how to compute a neighboring singular Lagrangian value in the direction of $B$. 
We accomplish this by storing auxiliary information as the DP algorithm (\ref{eq:L_DP}) is computed for a fixed multiplier $\lambda$, in order to identify a neighboring optimal solution to (\ref{eq:DualProblem}) if $\lambda$ is increased / decreased appropriately.

Specifically, we compute a neighboring singular multiplier value within the same DP framework (\ref{eq:L_DP}) developed to solve (\ref{eq:DualProblem}) as follows.
Denote by $(v_{n+1}^*, q_{v_{n+1}}^*)$ the argument that minimizes the sub-problem $\Phi_{v_n}(q_{v_n})$ in (\ref{eq:L_DP}) for a given $\lambda$. Further, denote by $\Psi_{v_n}(q_{v_n})$ and $\Upsilon_{v_n}(q_{v_n})$ the distortion and rate of the optimal solution $(v_{n+1}^*, q_{v_{n+1}}^*)$ to sub-problem $\Phi_{v_n}(q_{v_n})$ respectively. $\Psi_{v_n}(q_{v_n})$ and $\Upsilon_{v_n}(q_{v_n})$ can be computed and stored in DP tables as (\ref{eq:L_DP}) is being solved recursively; specifically, they are computed as:
\begin{align}
\label{eq:Psi}
\Psi_{v_n}(q_{v_n}) & = \Delta_{v_{n+1}^*}(v_n, q_{v_n}, q_{v_{n+1}^*}) + \Psi_{v_{n+1}^*}(q_{v_{n+1}^*}) \\
\label{eq:Upsilon}
\Upsilon_{v_n} (q_{v_n}) & = r_{v_{n+1}^*}(v_n, q_{v_n}, q_{v_{n+1}^*}) + \Upsilon_{v_{n+1}^*}( q_{v_{n+1}^*}) 
\end{align}

\subsubsection{Computing a smaller Singular Value}

Suppose first that $R(\mathcal{V}_{\lambda}, \mathbf{q}_{\lambda}) < B$, and we need to decrease $\lambda$ in order to increase $R(\mathcal{V}_{\lambda}, \mathbf{q}_{\lambda})$. 
To find the neighboring \textit{smaller} singular value $\lambda^-$, where $\lambda^- < \lambda$, we know that $\lambda^-$ and $\lambda$ share an optimal solution $(\mathcal{V}_{\lambda}, \mathbf{q}_{\lambda})$, and that $\lambda^-$ has an additional solution with rate \textit{larger} than $R(\mathcal{V}_{\lambda}, \mathbf{q}_{\lambda})$. 
This additional globally optimal solution $(\mathcal{V}_{\lambda^-}, \mathbf{q}_{\lambda^-})$ must contain a new local solution of a sub-problem $\Phi_{v_n}(q_{v_n})$ as $\lambda$ decreases. 
Thus, we seek the closest multiplier $\lambda^-$ to $\lambda$, $\lambda^- < \lambda$, where there exists a sub-problem $\Phi_{v_n}(q_{v_n})$ with a new local solution $(v_{n+1}^-, q_{v_{n+1}}^-)$ whose rate $R(v_{n+1}^-, q_{v_{n+1}}^-)$ is larger than the previous solution's $R(v_{n+1}^*, q_{v_{n+1}}^*)$.

In particular, for each sub-problem $\Phi_{v_n}(q_{v_n})$ we compute a \textit{singular value candidate} $\lambda^-_{v_n}(q_{v_n})$, where $\lambda^-_{v_n}(q_{v_n}) < \lambda$, as:
\begin{equation}
\label{eq:lambda1}
\lambda^-_{v_n}(q_{v_n}) = 
\underset{\substack{v \in \mathcal{V}^o  \\ v > v_n \\ q_{v} \in \mathcal{Q}}} {\mathrm{max}} \frac{\Psi_{v_{n}}(q_{v_{n}}) -(\Delta_{v}(v_{n}, q_{v_{n}}, q_{v}) + \Psi_{v}(q_{v}) )}{(r_{v}(v_{n}, q_{v_n}, q_{v}) + \Upsilon_{v}(q_{v})) - \Upsilon_{v_n}(q_{v_n})}
\end{equation}
where the search for the maximization is performed over the set of units and QPs, $(v, q_{v})$, with a \textit{larger} resulting rate:
\begin{equation}
\label{eq:lambda2_search}
r_{v}(v_{n}, q_{v_n}, q_{v}) + \Upsilon_{v}(q_{v}) > \Upsilon_{v_n}(q_{v_n})
\end{equation}
 
In other words, $\lambda^-_{v_n}(q_{v_n})$ is the closest multiplier value smaller than $\lambda$ where the sub-problem $\Phi_{v_n}(q_{v_n})$ results in a different solution with a larger rate.
Geometrically, (\ref{eq:lambda1}) is computing an RD point on the convex hull to the right of $(v_{n+1}^*, q_{v_{n+1}}^*)$ with a larger rate. See Fig. \ref{fig:R_lambda}. 

\begin{figure}[b]
\centering
\includegraphics[width=2.95in]{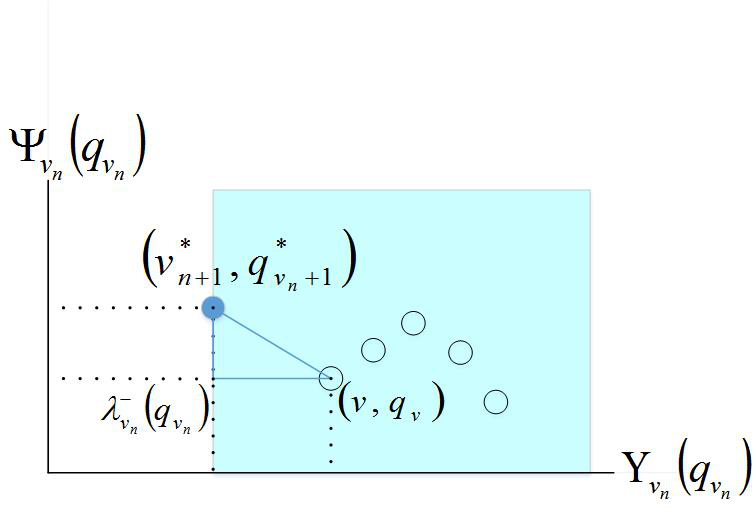}
\vspace{-0.1in}
\caption{Illustration of the search of singular value candidate $\lambda^-_{v_n}(q_{v_n})$ for the $\Phi_{v_n}(q_{v_n})$ sub-problem. The search is done over the points $(v, q_v)$ with larger rates than current optimal solution $(v^*_{n+1}, q^*_{v_{n+1}})$ (blue region). } 
\label{fig:R_lambda}
\end{figure}

The emergence of a new globally optimal solution $(\mathcal{V}_{\lambda^-}, \mathbf{q}_{\lambda^-})$ can stem from any sub-problem $\Phi_{v_n}(q_{v_n})$ as $\lambda$ decreases. To identify the first sub-problem $\Phi_{v_n}(q_{v_n})$ that results in a new local solution, we compute $\lambda^-$ as the \textit{largest} (closest to $\lambda$) of all singular value candidates $\lambda^-_{v_n}(q_{v_n})$:
\begin{equation}
\label{eq:Pi_1}
\lambda^- = \underset{v_n \in \mathcal{V}^o, q_{v_n} \in \mathcal{Q}} {\mathrm{max}} \; 
\lambda^-_{v_n}(q_{v_n})
\end{equation}
Denote by $(v_n^-, q_{v_n}^-)$ the argument that maximizes (\ref{eq:Pi_1}). 
Further, denote by $(v_{n+1}^-, q_{v_{n+1}}^-)$ the argument that maximizes (\ref{eq:lambda1}) for sub-problem $\Phi_{v_n^-}(q_{v_n}^-)$.
We show that using $\lambda^-$, sub-problem $\Phi_{v_n^-}(q_{v_n}^-)$ has two simultaneously optimal solutions.

\vspace{0.05in}
\begin{lemma}
Using singular value $\lambda^-$, sub-problem $\Phi_{v_n^-}(q_{v_n}^-)$ has two solutions, $(v_{n+1}^*, q_{v_{n+1}}^*)$ and $(v_{n+1}^-, q_{v_{n+1}}^-)$, that are simultaneously optimal. 
\end{lemma}

\vspace{0.05in}
\begin{proof}
First, because $\lambda^-$ is the largest among all singular value candidates, when multiplier is $\lambda^-$, sub-problems other than $\Phi_{v_n^-}(q_{v_n}^-)$ still have the same optimal solutions as before when multiplier was $\lambda$. 
This means that except for $(v_n^-, q_{v_n}^-)$, distortions $\Psi_{v_n}(q_{v_n})$ and rates $\Upsilon_{v_n}(q_{v_n})$ for all sub-problems remain the same when multiplier is $\lambda^-$.
Consider now sub-problem $\Phi_{v_n^-}(q_{v_n}^-)$.
For candidates $(v, q_v)$ with rates smaller than $(v_{n+1}^*, q_{v_{n+1}}^*)$, optimality of $(v_{n+1}^*, q_{v_{n+1}}^*)$ when multiplier is $\lambda$ means that the RD cost of $(v_{n+1}^*, q_{v_{n+1}}^*)$ remains smaller than these candidates $(v, q_v)$ when multiplier is now smaller.
One can then show that the definition $\lambda^-_{v_n^-}(q_{v_n^-})$ in (\ref{eq:lambda1}) implies that RD costs of $(v_{n+1}^*, q_{v_{n+1}}^*)$ and $(v_{n+1}^-, q_{v_{n+1}}^-)$ are the same.
Finally, because $(v_{n+1}^*, q_{v_{n+1}}^*)$ and $(v_{n+1}^-, q_{v_{n+1}}^-)$ are neighboring convex-hull points, candidates $(v, q_v)$ different from $(v_{n+1}^-, q_{v_{n+1}}^-)$ with rates larger than $(v_{n+1}^*, q_{v_{n+1}}^*)$, must result in a larger RD cost than $(v_{n+1}^*, q_{v_{n+1}}^*)$ and $(v_{n+1}^-, q_{v_{n+1}}^-)$ when multiplier is $\lambda^-$.
\end{proof}

\vspace{0.05in}
We note that the computed $\lambda^-$ in (\ref{eq:Pi_1}) only guarantees that a new local solution has emerged from a sub-problem $\Phi_{v_n}(q_{v_n})$. The globally optimal solution will not change \textit{if} the changed sub-problem $\Phi_{v_n}(q_{v_n})$ is not part of the global solution. However, successive moves to $\lambda^-$ will eventually trigger a change in the global solution, resulting in a new rate $R(\mathcal{V}_\lambda, \mathbf{q}_\lambda)$.

\subsubsection{Computing a larger Singular Value}

Similarly, we can compute the neighboring singular value larger than $\lambda$. The singular value candidate $\lambda^+_{v_n}(q_{v_n})$ larger than $\lambda$ for each sub-problem $\Phi_{v_n}(q_{v_n})$ is computed as:
\begin{equation}
\label{eq:lambda2}
\lambda^+_{v_n}(q_{v_n}) = 
\underset{\substack{v \in \mathcal{V}^o  \\ v > v_n  \\ q_{v} \in \mathcal{Q}}} {\mathrm{min}} \frac{\left (\Delta_{v}(v_{n}, q_{v_{n}}, q_{v}) + \Psi_{v}(q_{v}) \right)  - \Psi_{v_{n}}(q_{v_{n}})}{\Upsilon_{v_{n}}(q_{v_{n}}) - (r_{v}(v_{n}, q_{v_{n}}, q_{v}) + \Upsilon_{v}(q_{v}))}
\end{equation}
where the search for the minimization is performed over the set of units $v$ and QPs $q_{v}$ with a \textit{smaller} resulting rate:
\begin{equation}
\label{eq:lambda2_search}
\Upsilon_{v_{n}}(q_{v_{n}}) > r_{v}(v_{n}, q_{v_{n}}, q_{v}) + \Upsilon_{v}(q_{v}) 
\end{equation}
 
Then, the singular value $\lambda^+$ is the \textit{smallest} of all singular value candidates $\lambda^+_{v_n}(q_{v_n})$:
\begin{equation}
\label{eq:Pi_2}
\lambda^+ = \underset{v_n \in \mathcal{V}^o, q_{v_n} \in \mathcal{Q}} {\mathrm{min}} \; 
\lambda^+_{v_n}(q_{v_n})
\end{equation}



\subsubsection{DP Table Update}

From (\ref{eq:Pi_1}) and (\ref{eq:Pi_2}), we know that, given an initial $\lambda$ value, the neighboring smaller ($\lambda^-$) or larger singular value ($\lambda^+$) can be found with their corresponding solutions using (\ref{eq:L_DP}). 
The decision of which direction we should march to find the next singular value is determined by the rate of the computed solution  $R(\mathcal{V}_\lambda, \mathbf{q}_\lambda)$. 
Then, we use (\ref{eq:L_DP}) to update entries in the DP table given new multiplier $\lambda^-$ or $\lambda^+$. 
Note that only a subset of DP table entries need to be updated. 
Specifically, since in (\ref{eq:L_DP}) only candidates $(v_{n+1}, q_{v_{n+1}})$ where $v_{n+1} > v_n$ are considered, given a new singular value $\lambda^-$ or $\lambda^+$ associated with sub-problem $\Phi_{v_n^-}(q_{v_n^-})$ or $\Phi_{v_n^+}(q_{v_n^+})$, only DP entries $(v, q)$, $v < v_n^-$ or $v < v_n^+$, require updates using (\ref{eq:L_DP}). This further reduces the complexity of our algorithm.



\alglanguage{pseudocode}
\begin{algorithm}[!b]
\caption{Search of the optimal Lagrange multiplier}\label{algo}
\begin{algorithmic}[1]
\State Initialize $\lambda$
\State \emph{Coarse-grained search:} Perform a binary search of $\lambda$ until no new solutions can be reached.
\State \emph{Fine-grained search:}
\State Solve (\ref{eq:DualProblem}) via (\ref{eq:L_DP}) with unique solution $R(\mathcal{V}_{\lambda}, \mathbf{q}_{\lambda})$.
\If {$R(\mathcal{V}_{\lambda}, \mathbf{q}_{\lambda}) < B$}
\State Find singular value $\lambda^-$ via (\ref{eq:Pi_1}), with $\lambda^- < \lambda$.
\State $ \lambda  \gets \lambda^-$
\Else
\State Find singular value $\lambda^+$ via (\ref{eq:Pi_2}), with $\lambda^+ > \lambda$.
\State $ \lambda  \gets \lambda^+$
\EndIf
\Repeat
\State Update the DP entries $(v_j, q_{v_j})$ that needs to be modified, $v_j < v$, when $v$ is associated to $\lambda$.
\State Find simultaneous solutions $(\mathcal{V}_{\lambda}^l, \mathbf{q}_{\lambda}^l)$ and $(\mathcal{V}_{\lambda}^u, \mathbf{q}_{\lambda}^u)$, where $R(\mathcal{V}_{\lambda}^l, \mathbf{q}_{\lambda}^l) < R(\mathcal{V}_{\lambda}^u, \mathbf{q}_{\lambda}^u)$.
\If {$R(\mathcal{V}_{\lambda}^u, \mathbf{q}_{\lambda}^u) < \bar{B}$}
\State Find singular value $\lambda^-$ via (\ref{eq:Pi_1}), with $\lambda^- < \lambda$.
\State $ \lambda  \gets \lambda^-$
\ElsIf{$\bar{B} < R(\mathcal{V}_{\lambda}^l, \mathbf{q}_{\lambda}^l)$}
\State Find singular value $\lambda^+$ via (\ref{eq:Pi_2}), with $\lambda^+ > \lambda$.
\State $ \lambda  \gets \lambda^+$
\EndIf
\Until{$R(\mathcal{V}_{\lambda}^l, \mathbf{q}_{\lambda}^l) \leq \bar{B} \leq R(\mathcal{V}_{\lambda}^u, \mathbf{q}_{\lambda}^u) $}
\State $(\mathcal{V}_{\lambda}^l, \mathbf{q}_{\lambda}^l)$ is the best approximate Lagrangian solution.
\end{algorithmic}
\end{algorithm}

\subsubsection{Complexity Analysis}

Computing (\ref{eq:lambda1}) or (\ref{eq:lambda2}) for each sub-problem $\Phi_{v_n^-}(q_{v_n^-})$ or $\Phi_{v_n^+}(q_{v_n^+})$ has complexity $O(V Q)$. There are $O(V Q)$ sub-problems, and hence computing $\lambda^-$ or $\lambda^+$ is $\mathcal{O}(V^2 Q^2)$. Denote by $m$ the number of iterations until the optimal singular multiplier value is found. 
Thus the multiplier search complexity is $\mathcal{O}(m V^2 Q^2)$.

The number of iterations depends on how far from the optimal multiplier is the initial $\lambda$ value. 
To reduce the number of iterations, we propose a hybrid coarse- / fine-grained multiplier search strategy. First, we perform a non-singular-value binary search, as done in \cite{shoham88}, to produce big changes in $\lambda$ to approach the optimal multiplier. When binary search fails to yield new solutions, we apply our fine-grained singular-value search until the optimal multiplier value is found.
The search strategy for the best multiplier $\lambda^*$ is summarized in Algorithm \ref{algo}.

\section{Experimental Results}
\label{sec:results}
\subsection{Experimental Setup}

We now demonstrate the performance of our proposed rate allocation algorithm (\ref{eq:L_DP}) with optimal Lagrange multiplier selection in monoview and multiview video coding problems. 
Each data unit represents a full frame in a monoview video or a view in a multiview video. 
We consider the monoview video datasets \emph{Hall Monitor} (352 $\times$ 288, 30fps) \cite{Yue2009,Yue2009_2} and  \emph{Kimono} (1920 $\times$ 1080, 24fps), provided by  Nakajima Laboratory of the Tokyo Institute of Technology. 
Both sequences have a GOP size of 1s, namely 30 frames and 24 frames, respectively. 
Our algorithm is used to select frames and corresponding QPs for coding in each GOP.

For the multiview video datasets, we use three sequences: \emph{Shark} (1920 $\times$ 1088, 30fps, 9 views), provided by NICT for MPEG FTV standardization \cite{shark}, \emph{Undo Dancer} (1920 $\times$ 1088, 25fps, 5 views) \cite{Dancer} and \emph{Soccer Linear2} (1600 $\times$ 1200, 60fps, 7 views) \cite{goorts2014}. 
We used a GOP size of 8 frames and an intra-period of 24 frames 
as defined under the common test conditions by JCT-3V \cite{karsten2014}. 
Our algorithm is used to select the views to be encoded and their corresponding QPs; hierarchical B-frames \cite{sullivan12} are used in the temporal dimension to exploit temporal redundancy as done in the standard. 
A cascading quantization parameter (CQP) strategy \cite{Liu2007} is used to assign the QPs to the frames in the GOPs, where a $\Delta$QP is added to the QP value of the anchor frame to generate the QP values of the various frames in the GOP. We consider a $\Delta$QP vector equal to $\{0, 1, 2, 3, 4, 4, 3, 4\}$ for the successive frames in the GOP, as suggested in the reference software of 3D-HEVC  \cite{htm}.



Since each view consists of a texture and a depth image pair, we use our algorithm to select image pairs for coding and QPs for the texture images only, while QPs for the depth images are fixed at 30, so that 3D geometry information are coded accurately for high-quality virtual view synthesis. 
While QPs for texture and depth images can be jointly selected for optimal RD performance as done in \cite{cheung11tip2} for static multiview image sequences, in practice depth images represent only a small fraction of the total bitrate (about 10\%), and we optimize only texture image QPs.
We note that extension of our optimization to include selection of multiple QPs for a given data unit is relatively straightforward and is left for future work.

In our experiments, we compute PSNR of the luminance component (Y-PSNR) for all the decoder-side data units that are decoded or interpolated if they are left uncoded at the encoder. In particular, to reconstruct uncoded frames in monoview video sequences we use a popular temporal up-sampling method based on motion estimation \cite{LarsLau12,zhai05}.
To construct missing views in multiview video sequences, we use a simple depth-image based rendering (DIBR) \cite{Schmeing2011,Zinger2010} method at the decoder where the color pixels from the closest right and left coded views are projected to the missing intermediate view given per-pixel disparity information provided by the corresponding depth images. 

For comparison, we evaluate the performance of the rate control (RC) schemes \cite{bin12,woong13} adopted by the reference software HM 15.0 \cite{hm} of the High Efficiency Video Coding (HEVC) standard \cite{sullivan12} for monoview video sequences, and by the reference software HTM 13.0  \cite{htm} of the 3D extension of HEVC (3D-HEVC) \cite{muller13} for multiview video. These solutions only optimize QPs of different frames and do not skip data units at the encoder. For fairness, we also fix QP of the depth images when RC schemes of the reference software for 3D-HEVC is evaluated, so that only QPs of the texture images of the multiview video sequences are optimized. 



In the following, the performance of our algorithm is illustrated for monoview and multiview video sequences in two scenarios: (i) when data units are independently coded and (ii) when data units are predictively coded.

\subsection{Independent Coding}

We first evaluate the performance of competing RC schemes for the case of independently encoded units for monoview and multiview video sequences. The available set of QPs for the coding units are $\mathcal{Q}=\{ 25, 26, \ldots, 51\}$ for both cases.

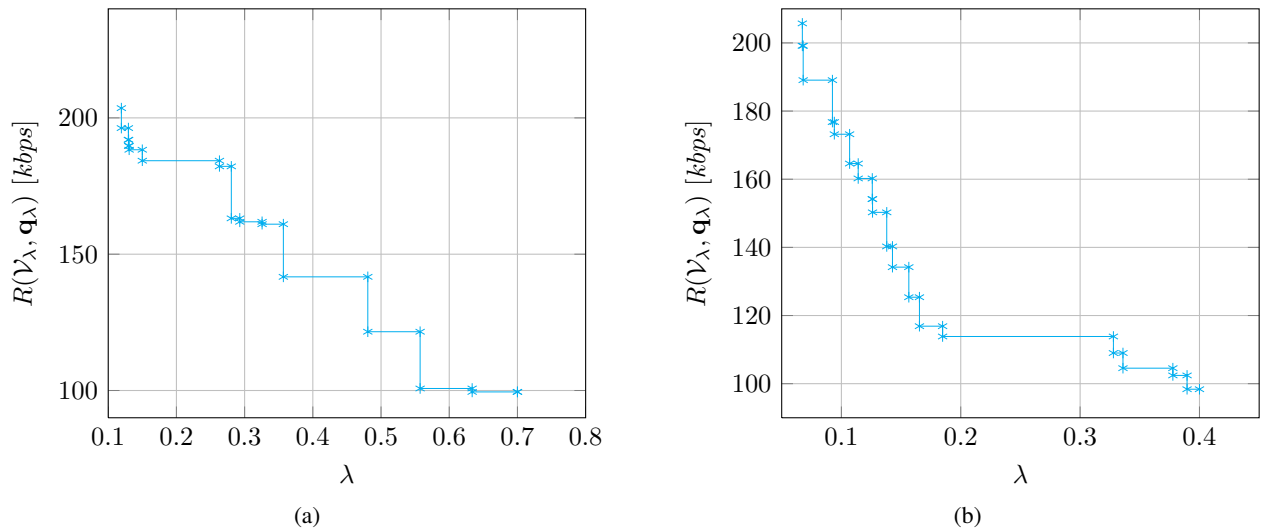
\begin{figure*}[!t]
\centering
	\subfloat[]{
	\begin{tikzpicture}
	        \tikzstyle{every node}=[font=\small]
	        \begin{axis}[%
	            width=0.35\textwidth,
	            height=0.3\textwidth,
	            at={(0\textwidth,0\textwidth)},
	            scale only axis,
	            xmin=0.1,
	            xmax=0.8,
	            xlabel={\small $\lambda$},
	            xmajorgrids,
	            ymin=90,
	            ymax=240,
	             ylabel={\small $R(\mathcal{V}_\lambda, \mathbf{q}_\lambda)$ $[kbps]$},
	            ymajorgrids,
	            cycle list name=color list,
	            legend style={at={(0.97,0.95)},anchor=north east,legend cell align=left,align=left,draw=white!15!black}
	            ]
	            \addplot [color=cyan,solid,mark=asterisk,mark options={solid},very thin]
	                table[row sep=crcr]{%
	        0.7000   99.4800\\
	        0.7000   99.4800 \\
   			0.6336   99.4800 \\
    			0.6336  100.7400 \\
    			0.5573  100.7400 \\
   			0.5573  121.5600 \\
    			0.4806  121.5600 \\
   			0.4806  141.6900 \\
    			0.3568  141.6900 \\
    			0.3568  161.0100 \\
    			0.3256  161.0100 \\
    			0.3256  161.8800 \\
    			0.2929  161.8800  \\
    			0.2929  163.1400  \\
    			0.2805  163.1400 \\
    			0.2805  182.2200 \\
    			0.2631  182.2200 \\
    			0.2631  184.2900 \\
    			0.1500  184.2900 \\
    			0.1500  188.3100 \\
    			0.1307  188.3100 \\
    			0.1307  189.7200 \\
    			0.1299  189.7200 \\
    			0.1299  191.9700 \\
    			0.1296  191.9700 \\
    			0.1296  196.2600 \\
    			0.1193  196.2600 \\
    			0.1193  203.5800\\
	            };
	        \end{axis}
	    \end{tikzpicture}
	\label{fig:R_Lambda_ind_Shark}
	}
	\hfil
	\subfloat[]{
	\begin{tikzpicture}
	        \tikzstyle{every node}=[font=\small]
	        \begin{axis}[%
	            width=0.35\textwidth,
	            height=0.3\textwidth,
	            at={(0\textwidth,0\textwidth)},
	            scale only axis,
	            xmin=0.05,
	            xmax=0.45,
	            xlabel={\small $\lambda$},
	            xmajorgrids,
	            ymin=90,
	            ymax=210,
	            ylabel={\small $R(\mathcal{V}_\lambda, \mathbf{q}_\lambda)$ $[kbps]$},
	            ymajorgrids,
	            cycle list name=color list,
	            legend style={at={(0.97,0.95)},anchor=north east,legend cell align=left,align=left,draw=white!15!black}
	            ]
	            \addplot [color=cyan,solid,mark=asterisk,mark options={solid},very thin]
	                table[row sep=crcr]{%
	        0.4      98.3750  \\
	        0.3896   98.3750  \\
	        0.3896   102.4250 \\
     0.3777  102.4250  \\
     0.3777  104.5500 \\
     0.3359  104.5500 \\
     0.3359  109.0000 \\
     0.3279  109.0000 \\
     0.3279  113.8500 \\
     0.1848  113.8500 \\
     0.1848  116.8750 \\
     0.1654  116.8750 \\
     0.1654  125.3750 \\
     0.1565  125.3750 \\
     0.1565  134.2000 \\
     0.1428  134.2000 \\
     0.1428  140.2750 \\
     0.1380  140.2750 \\
     0.1380  150.2750 \\
     0.1261  150.2750 \\
     0.1261  154.1750 \\
     0.1259  154.1750 \\
     0.1259  160.2250 \\
     0.1141  160.2250 \\
     0.1141  164.6000 \\
     0.1069  164.6000 \\
     0.1069  173.2000 \\
     0.0941  173.2000 \\
     0.0941  176.7750 \\
     0.0925  176.7750 \\
     0.0925  189.0750 \\
     0.0680  189.0750 \\
     0.0680  199.1500 \\
     0.0673  199.1500 \\
     0.0673  205.7250 \\
	            };
	        \end{axis}
	    \end{tikzpicture}
	\label{fig:R_Lambda_ind_UndoDancer}
	}
	\caption{Relationship between the rate $R(\mathcal{V}_\lambda, \mathbf{q}_\lambda)$ and the Lagrange multiplier $\lambda$ for the (a) \emph{Shark} and (b) \emph{Undo Dancer} multiview video sequences when views are independently encoded. }
	\label{fig:R_Lambda_ind}
	\end{figure*}

To examine the behaviour of our algorithm, we show in Fig. \ref{fig:R_Lambda_ind} the relationship between the rate $R(\mathcal{V}_\lambda, \mathbf{q}_\lambda)$ of optimal Lagrangian solutions and Lagrange multiplier $\lambda$ for the multiview sequences \emph{Shark} and \emph{Undo Dancer}. 
In particular, we illustrate the iterative multiplier search process to identify, among all Lagrangian solutions to (\ref{eq:DualProblem}) for any $\lambda$, one that minimizes the aggregate distortion subject to a rate budget, which in this case is $B = 200 kbps$. 
Multiplier $\lambda$ is initialized to be  $\lambda=0.7$ and $\lambda=0.4$ for for the two sequences \emph{Shark} and \emph{Undo Dancer} respectively. 
The optimal solution $(\mathcal{V}_\lambda, \mathbf{q}_\lambda)$ is reached in these two cases at singular value $\lambda=0.1193$ for \emph{Shark} $\left(R(\mathcal{V}_\lambda, \mathbf{q}_\lambda) = 196.26\right)$ and at singular value $\lambda=0.0673$ for \emph{Undo Dancer} $\left(R(\mathcal{V}_\lambda, \mathbf{q}_\lambda) = 199.15\right)$.
Using (\ref{eq:bound}), the distortion bound for each sequence can be computed using the two simultaneously optimal solutions at the optimal singular value, which is 0.01 dB and 0.03 dB for \emph{Shark} and \emph{Undo Dancer}, respectively.
This shows in practice that the $R(\mathcal{V}_\lambda, \mathbf{q}_\lambda)$ versus multiplier $\lambda$ plot is dense with samples (\textit{i.e.}, the RD plots of $\Phi_{v_n}(q_{v_n})$ are in general convex), and the resulting distortion bounds are tight.

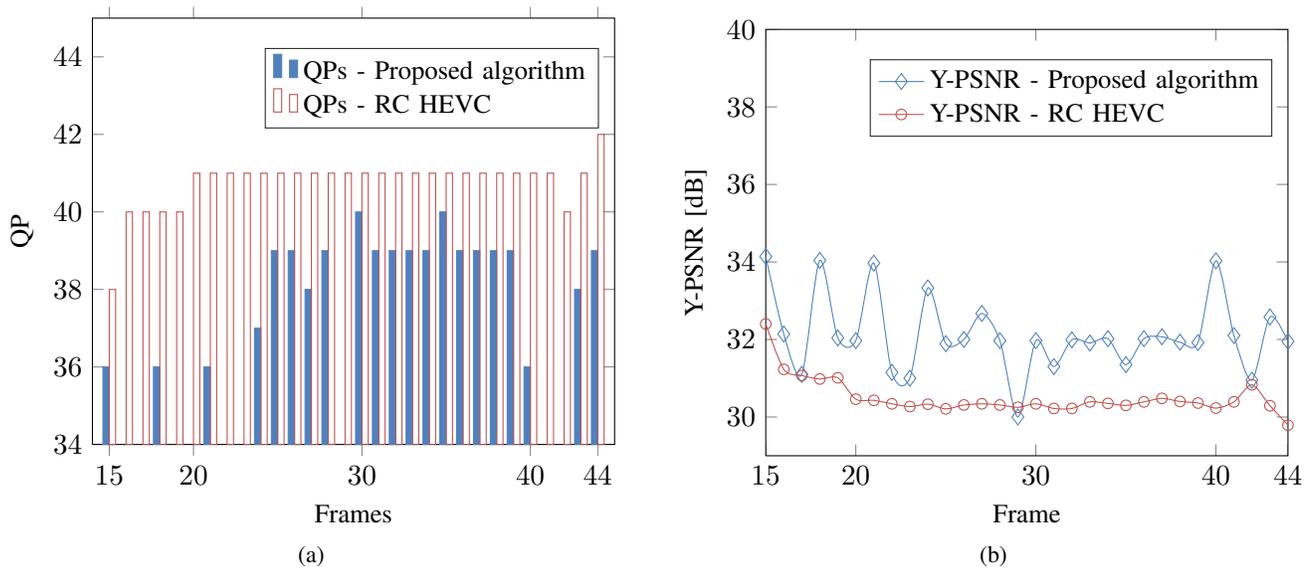
\begin{figure*}{}
\subfloat[]{\begin{tikzpicture}
\begin{axis}[
legend style={at={(0.97,0.93)},anchor=north east,legend cell align=left,align=left,draw=white!15!black},
	x tick label style={
		/pgf/number format/1000 sep=},
	ylabel={\small QP},
	ybar=0pt,
	bar width=2.4pt,
	ymin=34, ymax=45,
	xmin=14, xmax=45,
	xtick={15,20,30,40,44},
	xlabel={\small Frames},
	height=0.4\textwidth,
	width=0.47\textwidth,
]
\addplot[bblue, fill=bblue]
	coordinates {(15,36) (18,36) (21,36) (24,37) (25,39) (26,39) (27,38) (28,39)
		 (30,40) (31,39) (32,39) (33,39) (34,39) (35,40) (36,39) (37,39) (38,39) (39,39) (40,36) 		    (43,38) (44,39)}; \addlegendentry{\small QPs - Proposed algorithm}
\addplot[rred,pattern color =rred,
    postaction={
        pattern=north east lines
    }
]  
	coordinates {(15,38) (16,40) (17,40) (18,40) (19,40) (20,41) (21,41) (22,41) (23,41) (24,41) (25,41) (26,41) (27,41) (28,41) (29,41) (30,41) (31,41) (32,41) (33,41) (34,41) (35,41) (36,41) (37,41) (38,41) (39,41) (40,41) (41,41) (42,40) (43,41) (44,42)}; \addlegendentry{\small QPs - RC HEVC}
\end{axis}
\end{tikzpicture}\label{fig:HallMonitor_Ind_QP}}
\hfil 
\subfloat[]{\begin{tikzpicture}
\begin{axis}[
   legend style={at={(0.97,0.93)},anchor=north east,legend cell align=left,align=left,draw=white!15!black},  
  ymin=29, ymax=40,
  xmin=15, xmax=44,
  xtick={15,20,30,40,44},
  ylabel={\small Y-PSNR [dB]},
  xlabel={\small Frame},
  height=0.4\textwidth,
  width=0.47\textwidth
 ]
\addplot[smooth,mark=diamond,mark options={solid},mark size=3pt, bblue]
  coordinates{(15,34.14) (16,32.14) (17,31.10) (18,34.04) (19,32.04) (20,31.97)  (21,33.97) (22,31.15) (23,31.00) (24,33.33) (25,31.89) (26,32.00) (27,32.67) (28,31.97) (29,30.00) (30,31.97) (31,31.30) (32,31.99) (33,31.91) (34,32.02) (35,31.35) (36,32.02) (37,32.07) (38,31.93) (39,31.92) (40,34.03) (41,32.10) (42,30.95) (43,32.58) (44,31.95)}; \addlegendentry{ \small Y-PSNR - Proposed algorithm}
  
\addplot[smooth,mark=o,mark options={solid},rred]
  coordinates{(15,32.40) (16,31.23) (17,31.07) (18,30.98) (19,31.01) (20,30.46)  (21,30.43) (22,30.34) (23,30.27) (24,30.33) (25,30.21) (26,30.31) (27,30.34) (28,30.31) (29,30.25) (30,30.34) (31,30.22) (32,30.22) (33,30.39) (34,30.35) (35,30.30) (36,30.39) (37,30.48) (38,30.40) (39,30.36) (40,30.23) (41,30.39) (42,30.83) (43,30.29) (44,29.79)}; \addlegendentry{ \small Y-PSNR - RC HEVC}  
  
\end{axis}
\end{tikzpicture}\label{fig:HallMonitor_Ind_PSNR}}
\caption{Frame-by-frame comparison of our proposed algorithm and the RC of HEVC for \emph{Hall Monitor} monoview video sequence: (a) QP selection and (b) quality comparison (Y-PSNR) for a rate budget $B=500 kbps$. $R=490.60 kbps$ and an average Y-PSNR=32.11 dB is achieved by our solution, and $R= 499.53 kbps$ and Y-PSNR=30.50 dB is achieved by the RC of HEVC.}\label{fig:HallMonitor_Ind}
\end{figure*}

To visualize a particular solution, we compare the QP selection for each selected data unit, obtained as the best solution of our algorithm, to the QP adaptation solution of the RC of HEVC. In Fig. \ref{fig:HallMonitor_Ind_QP}, we show the solutions for one GOP (from frame 15 to 44) of the \emph{Hall Monitor} monoview sequence. For these results, we consider a rate budget of 500kbps, and our algorithm achieves a rate of 490.60kbps and the RC of HEVC a rate of 499.53kbps. 
Note that $QP=0$ means that the frame is skipped and needs to be reconstructed at the decoder. 
Most of the frames that are skipped with our algorithm are the frames between frames 15 and 23, which correspond to the part with the lowest motion in the GOP under consideration. 
In addition, it can be seen that our algorithm assigns low QP values to the frames that are neighbors of the dropped frames, since these neighbors are used as reference frames in their reconstruction at the decoder.  
In Fig. \ref{fig:HallMonitor_Ind_PSNR}, we show Y-PSNR of coded ($QP > 0$) and interpolated ($QP = 0$) frames, for the same experiment as in Fig. \ref{fig:HallMonitor_Ind_QP}. 
We see that the solution of our algorithm achieves a higher average quality compared to the solution of the RC of HEVC, achieving an average Y-PSNR of $32.11dB$, while the RC solution has an average Y-PSNR of $30.50dB$. 


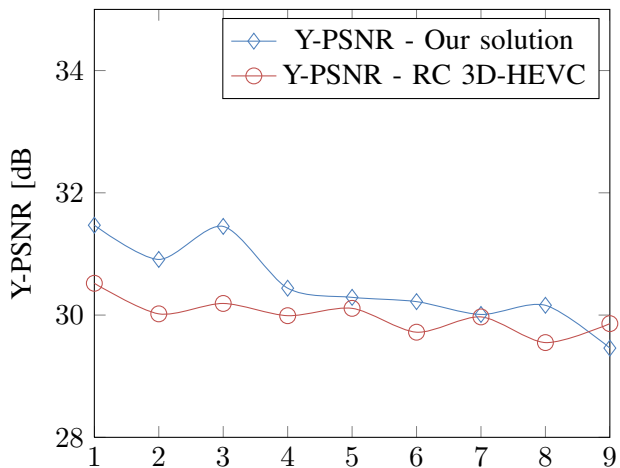
\begin{figure}[!t]
\centering
\begin{tikzpicture}
\begin{axis}[
  xmin=1, xmax=9,
  ymin=28, ymax=35,
  xtick = {1,2,3,4,5,6,7,8,9},
  ylabel=Y-PSNR [dB],
  xlabel=views
]
\addplot[smooth,mark=diamond,mark options={solid},mark size=3pt, bblue]
  coordinates{(1,31.47) (2,30.91) (3,31.45) (4,30.44) (5,30.29) (6,30.22)  (7,30.01) (8,30.16) (9,29.46)}; \addlegendentry{Y-PSNR - Our solution} 
\addplot[smooth,mark=o,mark options={solid},mark size=3pt,rred]
  coordinates{(1,30.52) (2,30.02) (3,30.19) (4,29.99) (5,30.11) (6,29.72)  (7,29.97) (8,29.55) (9,29.86)}; \addlegendentry{Y-PSNR - RC 3D-HEVC}  
\end{axis}
\end{tikzpicture}
\caption{View-by-view average quality comparison (Y-PSNR) of our proposed algorithm and the RC of 3D-HEVC for \emph{Shark} multiview video sequence.and a rate budget $B=250 kbps$. $ R=245.49 kbps$ 
and an average Y-PSNR=30.49 dB is achieved by our solution, and $R= 242.55 kbps$ and Y-PSNR=29.98 dB is achieved by the RC of 3D-HEVC.}\label{fig:Shark_Ind_PSNR}
\end{figure}

A view-by-view quality evaluation that results from the QP selection of our algorithm and the RC of 3D-HEVC is illustrated in Fig. \ref{fig:Shark_Ind_PSNR}. We consider the \emph{Shark} multiview video sequence and a rate budget of 250 kbps, where the consumed rate of our algorithm is 245.49 kbps, while the RC of 3D-HEVC uses 242.55 kbps. Our algorithm solution in this cases does not skip any view at the encoder, and still it achieves a higher average Y-PSNR (30.49 dB) compared to the RC of 3D-HEVC (29.98 dB). Proving that, the good performance of our algorithm does not reside on its capability of skipping data units at the encoder.

\begin{figure*}
\centering
\subfloat[Frame 15 - Proposed algorithm - QP= 43]{\includegraphics[width=2.5in]{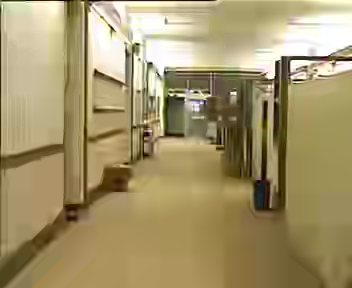}\label{subfig:HM_PA_F15}}\hfil
\subfloat[Frame 15 - RC HEVC - QP= 47]{\includegraphics[width=2.5in]{I_rec_frame_15_QP_47.png}\label{subfig:HM_RC_F15}}\\
\subfloat[Frame 17 - Proposed algorithm - Reconstructed frame]{\includegraphics[width=2.4in]{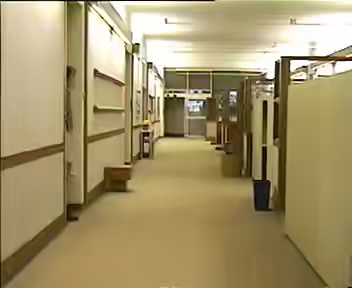}\label{subfig:HM_PA_F17}}\hfil
\subfloat[Frame 17 - RC HEVC - QP= 51]{\includegraphics[width=2.5in]{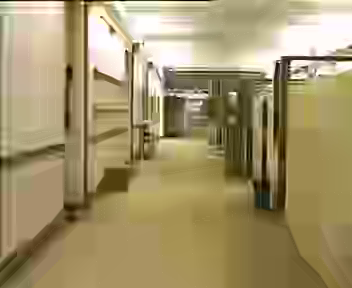}\label{subfig:HM_RC_F17}}
\caption{ Visual quality illustration for the \emph{Hall Monitor} monoview video sequence with independently encoded frames when the proposed algorithm and the RC of HEVC are used ($B=150$ kbps). (a) and (b) Show frame 15 encoded according to our proposed algorithm and the RC of HEVC, respectively. (c) Shows frame 17, that has has been skipped at the encoder and reconstructed at the decoder according to the proposed algorithm, achieving a higher visual quality compared to the RC of HEVC output in (d).}\label{chap2fig:HM_ind}
\end{figure*}

Tables \ref{tab:HallMonitor_Ind} and \ref{tab:Kimono_Ind} show the performance of both our rate allocation algorithm and the RC of HEVC, in terms of average Y-PSNR given a rate budget $B$ for the monoview sequences \emph{Hall Monitor} and \emph{Kimono}. 
We see that our algorithm always outputs a solution with a rate that is under the rate budget $B$ and achieves a higher quality, with a Y-PSNR gain of up to $2.34 dB$. The corresponding visual quality is illustrated in Fig. \ref{chap2fig:HM_ind} for \emph{Hall Monitor} for frames 15 and 17, when the rate budget is $B=150$ kbps. 
Our algorithm tends to skip frames with low motion, as frame 17, which are then efficiently interpolated at the decoder. 
This permits to achieve a higher visual quality than the one of the RC of HEVC that has to use a higher QP value to satisfy the rate budget.  

\begin{table}[!h]
\renewcommand{\arraystretch}{1.3}
\caption{Rate budget $B$, actual rate $R$ and average Y-PSNR value for the proposed algorithm and for the RC of HEVC, for the \emph{Hall Monitor} monoview video sequence with independently encoded frames.}
\centering
\begin{tabular}{|c|c c|c c|}
\hline
$B$ & \multicolumn{2}{c|}{Proposed algorithm} & \multicolumn{2}{c|}{RC HEVC} \\
$[kbps]$ & R $[kbps]$ & Y-PSNR $[dB]$ & R $[kbps]$ & Y-PSNR $[dB]$\\
\hline
150 & 149.39 & 26.62 & 149.96 & 24.93 \\
200 & 198.13 & 28.60 & 200.16 & 25.26 \\
300 & 298.88 & 29.57 & 300.31 & 27.29 \\
400 & 366.10 & 30.66 & 400.09 & 28.98 \\
500 & 490.60 & 30.79 & 499.53 & 30.49 \\
\hline
\end{tabular}
\label{tab:HallMonitor_Ind}
\end{table}

\begin{table}[!h]
\renewcommand{\arraystretch}{1.3}
\caption{Rate budget $B$, actual rate $R$ and average Y-PSNR value for the proposed algorithm and for the RC of HEVC, for the \emph{Kimono} monoview video sequence with independently encoded frames.}
\centering
\begin{tabular}{|c|c c|c c|}
\hline
B & \multicolumn{2}{c|}{Proposed algorithm} & \multicolumn{2}{c|}{RC HEVC} \\
$[kbps]$ & R $[kbps]$ & Y-PSNR $[dB]$ & R $[kbps]$ & Y-PSNR $[dB]$\\
\hline
150 & 149.03 & 27.44 & 149.81 & 26.42 \\
200 & 198.80 & 28.59 & 200.46 & 27.20 \\
300 & 296.32 & 29.56 & 299.97 & 28.37 \\
400 & 391.20 & 30.03 & 400.36 & 29.35 \\
500 & 499.61 & 30.23 & 502.06 & 30.17 \\
\hline
\end{tabular}
\label{tab:Kimono_Ind}
\end{table}

Similarly, Tables \ref{tab:Shark_Ind}, \ref{tab:UndoDancer_Ind} and \ref{tab:Soccer_Ind} show the performance of our algorithm and the RC of 3D-HEVC in terms of average Y-PSNR, with different rate budgets for the multiview sequences \emph{Shark}, \emph{Undo Dancer} and \emph{Soccer Linear2}. Although our algorithm has a better performance than the RC of 3D-HEVC in all cases, this gain is smaller than the monoview sequences. 
The main reason is that the frame-to-frame differences along the temporal dimension in monoview videos are relatively smaller, making skipping frames a more attractive option. 


\begin{table}[!h]
\renewcommand{\arraystretch}{1.3}
\caption{Rate budget $B$, actual rate $R$ and average Y-PSNR value for the proposed algorithm and for the RC of 3D-HEVC, for the \emph{Shark} multiview video sequence with independently encoded views.}
\centering
\begin{tabular}{|c|c c|c c|}
\hline
B & \multicolumn{2}{c|}{Proposed algorithm} & \multicolumn{2}{c|}{RC HEVC} \\
$[kbps]$ & R $[kbps]$ & Y-PSNR $[dB]$ & R $[kbps]$ & Y-PSNR $[dB]$\\
\hline
175 & 174.41 & 29.25 & 162.45 & 28.22 \\
200 & 196.26 & 29.26 & 198.75 & 29.22 \\
300 & 287.61 & 31.08 & 287.49 & 30.57 \\
400 & 381.54 & 32.35 & 389.28 & 31.51 \\
500 & 481.71 & 33.49 & 485.13 & 32.22 \\
\hline
\end{tabular}
\label{tab:Shark_Ind}
\end{table}

\begin{table}[!h]
\renewcommand{\arraystretch}{1.3}
\caption{Rate budget $B$, actual rate $R$ and average Y-PSNR value for the proposed algorithm and for the RC of 3D-HEVC, for the \emph{Undo Dancer} multiview video sequence with independently encoded views.}
\centering
\begin{tabular}{|c|c c|c c|}
\hline
B & \multicolumn{2}{c|}{Proposed algorithm} & \multicolumn{2}{c|}{RC HEVC} \\
$[kbps]$ & R $[kbps]$ & Y-PSNR $[dB]$ & R $[kbps]$ & Y-PSNR $[dB]$\\
\hline
175 & 169.15 & 27.71 & 169.55 & 26.99 \\
200 & 199.15 & 28.39 & 197.32 & 27.51 \\
300 & 280.55 & 29.57 & 286.95 & 28.74 \\
400 & 383.05 & 30.79 & 380.70 & 29.69 \\
500 & 476.78 & 31.63 & 486.15 & 30.47 \\
\hline
\end{tabular}
\label{tab:UndoDancer_Ind}
\end{table}

\begin{table}[!h]
\renewcommand{\arraystretch}{1.3}
\caption{Rate budget $B$, actual rate $R$ and average Y-PSNR value for the proposed algorithm and for the RC of 3D-HEVC, for the \emph{Soccer Linear2} multiview video sequence with independently encoded views.}
\centering
\begin{tabular}{|c|c c|c c|}
\hline
B & \multicolumn{2}{c|}{Proposed algorithm} & \multicolumn{2}{c|}{RC HEVC} \\
$[kbps]$ & R $[kbps]$ & Y-PSNR $[dB]$ & R $[kbps]$ & Y-PSNR $[dB]$\\
\hline
175 & 172.90 & 28.15 & 173.28 & 27.70 \\
250 & 245.98 & 28.89 & 249.06 & 28.99 \\
300 & 296.23 & 29.44 & 288.82 & 29.24 \\
400 & 395.32 & 30.30 & 398.11 & 30.77 \\
500 & 487.4 & 31.42 & 482.24 & 31.03 \\
\hline
\end{tabular}
\label{tab:Soccer_Ind}
\end{table}



\subsection{Predictive Coding}

We consider now the predictive coding case, and in particular, an $IPP \ldots$ predictive coding structure with one intra (I) coded data unit and subsequent predictively (P) coded units in each GOP. 
As the computational complexity in the rate allocation increases for predictive coding 
compared to the independent coding case---rate and distortion terms now depend on previous coded units---we decrease the granularity of the available QPs in the search space of our algorithm to $\mathcal{Q}=\{ 25, 28, 31 \cdots, 51\}$. 

For different rate constraints $B$, Tables \ref{tab:HallMonitor_Pred} and \ref{tab:Kimono_Pred} show the RD performance of our algorithm and the RC of HEVC for the \emph{Hall Monitor} and \emph{Kimono} monoview sequences. 
We observe that our assumption that the rate and distortion functions depend only on one reference frame (see Section \ref{sec:SM}) sometimes leads to under-estimation of the coding rate. Thus, we employ a simple post-processing step, where the Lagrangian solution to (\ref{eq:L_DP}) of a singular muliplier value $\lambda$ closest to the budget $B$ with actual aggregate coding rate below the budget is used.

Compared to the independent coding case, in the predictive coding scenario, the RC of HEVC does not lead to a good performance for the two monoview video sequences considered. This is evident for the \emph{Kimono} dataset in Table \ref{tab:Kimono_Pred} where the rates of some solutions of the RC of HEVC are far above the rate budget constraint (indicated by an asterisk), particularly at low bit budgets. 

We see from Tables \ref{tab:HallMonitor_Pred} and \ref{tab:Kimono_Pred}, that our algorithm typically achieves a better average quality. Moreover, our algorithm uses a sparser set of QPs than the RC of HEVC, meaning that our results can be improved if the same set of QPs $\mathcal{Q}$ is used by both schemes. 
Although our algorithm shows good performance on average, the gain of our algorithm is smaller for predictive coding than for independent coding. 
This is due to the fact that in predictive coding skipped frames have higher impact in the overall quality, which is one of the reasons of quality gains for independently encoded units. 
Indeed, when frames are skipped at the encoder during predictive coding, the distance between a coded frame and its reference increases, which reduces the coding performance. 
However, these results show that the good performance of our algorithm does not uniquely depend on its capability for skipping data units at the encoder.

\begin{table}[!h]
\renewcommand{\arraystretch}{1.3}
\caption{Rate budget $B$, actual rate $R$ and average Y-PSNR value for the proposed algorithm and for the RC of HEVC, for the \emph{Hall Monitor} monoview video sequence with predictively coded frames.}
\centering
\begin{tabular}{|c|c c|c c|}
\hline
B & \multicolumn{2}{c|}{Proposed algorithm} & \multicolumn{2}{c|}{RC HEVC} \\
$[kbps]$ & R $[kbps]$ & Y-PSNR $[dB]$ & R $[kbps]$ & Y-PSNR $[dB]$\\
\hline
50 & 47.96 & 30.06 & 50.14 & 29.92 \\
75 & 74.82 & 33.92 & 75.06 & 33.70 \\
100 & 96.84 & 34.17 & 99.89 & 34.28 \\
150 & 148.16 & 36.26 & 150.09 & 36.02 \\
\hline
\end{tabular}
\label{tab:HallMonitor_Pred}
\end{table}

\begin{table}[!h]
\renewcommand{\arraystretch}{1.3}
\caption{Rate budget $B$, actual rate $R$ and average Y-PSNR value for the proposed algorithm and for the RC of HEVC, for the \emph{Kimono} monoview video sequence with predictively coded frames.}
\centering
\begin{tabular}{|c|c c|c c|}
\hline
B & \multicolumn{2}{c|}{Proposed algorithm} & \multicolumn{2}{c|}{RC HEVC} \\
$[kbps]$ & R $[kbps]$ & Y-PSNR $[dB]$ & R $[kbps]$ & Y-PSNR $[dB]$\\
\hline
75 & 73.01 & 27.21 & 96.44 * & 28.01 \\
100 & 97.06 & 28.40 & 120.00 * & 29.02 \\
150 & 148.34 & 29.75 & 152.44 & 29.54 \\
200 & 186.44 & 30.41 & 204.12 & 30.24 \\
\hline
\end{tabular}
\label{tab:Kimono_Pred}
\end{table}

Finally, Tables \ref{tab:Shark_Pred}, \ref{tab:UndoDancer_Pred} and \ref{tab:Soccer_Pred} show the performance in terms of rate and average quality of our algorithm and the RC of 3D-HEVC for the predictive coding of the \emph{Shark}, \emph{Undo Dancer} and \emph{Soccer Linear2} multiview video sequences. 
Compared to the monoview video results presented in Tables \ref{tab:HallMonitor_Pred} and \ref{tab:Kimono_Pred} where the average gain is lower than 1dB, for predictively coded multiview video sequences better performance is observed, with an average quality gain higher than 1dB. 
Moreover, the solution obtained with our algorithm satisfies the rate budget most of the time with our original algorithm, and there is usually no need to use the post-processing step to modify the obtained solutions. 
This is due to the length of the prediction paths. 
In the case of multiview video, the maximum length of the (inter-view) prediction path is 9 views (\textit{e.g.}, \emph{Shark}), compared to 30 and 24 frames (GOP size) in the \emph{Hall Monitor} and \emph{Kimono} monoview video sequences. 
This means that, for the multiview video case, the effect of previously coded units in a current predicted unit is much more limited than in monoview video cases, thus making our assumption of Section \ref{sec:SM} more reasonable. 
In general, from these results we can conclude that when our algorithm is close to the rate budget (\textit{i.e.}, the granularity of the available QPs is not affecting the solution) it achieves a quality that is generally higher than the one in the RC of 3D-HEVC. 

%

\begin{table}[!h]
\renewcommand{\arraystretch}{1.3}
\caption{Rate budget $B$, actual rate $R$ and average Y-PSNR value for the proposed algorithm and for the RC of HEVC, for the \emph{Shark} multiview video sequence with predictively coded views.}
\centering
\begin{tabular}{|c|c c|c c|}
\hline
B & \multicolumn{2}{c|}{Proposed algorithm} & \multicolumn{2}{c|}{RC HEVC}  \\
$[kbps]$ & R $[kbps]$ & Y-PSNR $[dB]$ & R $[kbps]$ & Y-PSNR $[dB]$ \\
\hline
100 & 98.19 & 29.15 & 107.64 & 28.03 \\
150 & 147.25 & 31.02 & 147.60 & 29.51 \\
200 & 195.47 & 32.68 & 190.59 & 30.69 \\
300 & 297.22 & 33.51 & 290.91 & 32.59 \\
\hline
\end{tabular}
\label{tab:Shark_Pred}
\end{table}


\begin{table}[!h]
\renewcommand{\arraystretch}{1.3}
\caption{Rate budget $B$, actual rate $R$ and average Y-PSNR value for the proposed algorithm and for the RC of HEVC, for the \emph{Undo Dancer} multiview video sequence with predictively coded views.}
\centering
\begin{tabular}{|c|c c|c c|}
\hline
B & \multicolumn{2}{c|}{Proposed algorithm} & \multicolumn{2}{c|}{RC HEVC}  \\
$[kbps]$ & R $[kbps]$ & Y-PSNR $[dB]$ & R $[kbps]$ & Y-PSNR $[dB]$ \\
\hline
50 & 49.93 & 25.87 & 46.85 & 25.32 \\
75 & 74.15 & 28.23 & 74.75 & 27.54  \\
100 & 76.72 & 28.29 & 82.22 & 27.96  \\
150 & 129.81 & 30.25 & 148.07 & 30.47 \\
200 & 191.9 & 32.12 & 194.00 & 31.50 \\
\hline
\end{tabular}
\label{tab:UndoDancer_Pred}
\end{table}

\begin{table}[!h]
\renewcommand{\arraystretch}{1.3}
\caption{Rate budget $B$, actual rate $R$ and average Y-PSNR value for the proposed algorithm and for the RC of HEVC, for the \emph{Soccer Linear2} multiview video sequence with predictively coded views.}
\centering
\begin{tabular}{|c|c c|c c|}
\hline
B & \multicolumn{2}{c|}{Proposed algorithm} & \multicolumn{2}{c|}{RC HEVC}  \\
$[kbps]$ & R $[kbps]$ & Y-PSNR $[dB]$ & R $[kbps]$ & Y-PSNR $[dB]$ \\
\hline
100 & 98.96 & 28.06 & 93.06 & 26.09 \\
200 & 180.48 & 30.77 & 184.08 & 29.72  \\
250 & 180.48 & 30.77 & 247.02 & 31.04  \\
300 & 296.27 & 33.27 & 289.02 & 32.18 \\
400 & 396.18 & 34.71 & 376.14 & 33.53 \\
\hline
\end{tabular}
\label{tab:Soccer_Pred}
\end{table}

\section{Conclusion}
\label{sec:conclude}
A new and general solution for the optimal selection of the Lagrange multiplier in Lagrangian-based rate allocation optimization problems has been presented in this paper. It has been integrated into a general Lagrangian-based dynamic programming algorithm to efficiently solve rate allocation problems in video communication applications. The potential of our new algorithm has been illustrated in representative compression problems with monoview and multiview video sequences, and rate control solutions adopted by the reference softwares, HEVC and 3D-HEVC, have been used to appreciate the rate distortion performance of our proposed algorithm.


\subsection{Proof of Optimality}\label{Ap:Proof_of_Optimality}

We prove here that, if an optimal solution $(\mathcal{V}_{\lambda^*}, \mathbf{q}_{\lambda^*})$ to the unconstrained Lagrangian problem corresponding to multiplier value $\lambda^*$ satisfies the rate constraint exactly, \textit{i.e.},
\begin{equation}
R(\mathcal{V}_{\lambda^*}, \mathbf{q}_{\lambda^*}) = B,
\end{equation}

then $(\mathcal{V}_{\lambda^*}, \mathbf{q}_{\lambda^*})$ is also the optimal solution to the original constrained problem. The optimality of $(\mathcal{V}_{\lambda^*}, \mathbf{q}_{\lambda^*})$ implies that:

\begin{small}
\begin{equation}
D(\mathcal{V}_{\lambda^*}, \mathbf{q}_{\lambda^*}) + \lambda^* R(\mathcal{V}_{\lambda^*}, \mathbf{q}_{\lambda^*}) \leq D(\mathcal{V}, \mathbf{q}) + \lambda^* R(\mathcal{V}, \mathbf{q}), ~~
\forall \mathcal{V}, \mathbf{q} \nonumber
\end{equation}
\end{small}
Rearranging the terms, we get:
\begin{align}
\lambda^* \left[ R(\mathcal{V}_{\lambda^*}, \mathbf{q}_{\lambda^*}) - 
R(\mathcal{V}, \mathbf{q}) \right] \leq D(\mathcal{V}, \mathbf{q}) - D(\mathcal{V}_{\lambda^*}, \mathbf{q}_{\lambda^*}) \nonumber \\
\lambda^* \left[ B - R(\mathcal{V}, \mathbf{q}) \right] \leq D(\mathcal{V}, \mathbf{q}) - D(\mathcal{V}_{\lambda^*}, \mathbf{q}_{\lambda^*}) \nonumber 
\end{align}

Now we restrict our solution space to a subspace $\mathcal{S}$ where $R(\mathcal{V}, \mathbf{q}) \leq B$. Then, 

\begin{small}
\begin{align}
0 \leq \lambda^* \left[ B - R(\mathcal{V}, \mathbf{q}) \right] & \leq D(\mathcal{V}, \mathbf{q}) - D(\mathcal{V}_{\lambda^*}, \mathbf{q}_{\lambda^*}), ~~
\forall (\mathcal{V}, \mathbf{q}) \in \mathcal{S} \nonumber \\
D(\mathcal{V}_{\lambda^*}, \mathbf{q}_{\lambda^*}) & \leq D(\mathcal{V}, \mathbf{q}), ~~ \forall (\mathcal{V}, \mathbf{q}) \in \mathcal{S} \nonumber
\end{align}
\end{small}\noindent
We can thus conclude $(\mathcal{V}_{\lambda^*}, \mathbf{q}_{\lambda^*})$ is an optimal solution to the original constrained problem as well. $\Box$

\subsection{Performance Bound}\label{Ap:Performance_Bound}

We prove here the performance bound given in Eq. (\ref{eq:bound}). Let $(\mathcal{V}_{\lambda_1}, \mathbf{q}_{\lambda_1})$ and $(\mathcal{V}_{\lambda_2}, \mathbf{q}_{\lambda_2})$ be two solutions of the problem in (\ref{eq:DualProblem}) using $\lambda_1$ and $\lambda_2$ with resulting rates:

\begin{equation}
R(\mathcal{V}_{\lambda_1}, \mathbf{q}_{\lambda_1}) < B < R(\mathcal{V}_{\lambda_2}, \mathbf{q}_{\lambda_2})
\end{equation}

We can derive a performance bound for the feasible solution $(\mathcal{V}_{\lambda_1}, \mathbf{q}_{\lambda_1})$ as follows. Denote by $(\mathcal{V}^*, \mathbf{q}^*)$ the optimal solution to the original constrained problem. By the optimality of the solution $(\mathcal{V}_{\lambda_2}, \mathbf{q}_{\lambda_2})$, we can write:

\begin{align}
0 \leq \lambda^* \left[ R(\mathcal{V}_{\lambda_2}, \mathbf{q}_{\lambda_2}) - R(\mathcal{V}^*, \mathbf{q}^*) \right] & \leq D(\mathcal{V}^*, \mathbf{q}^*) - D(\mathcal{V}_{\lambda_2}, \mathbf{q}_{\lambda_2}) \nonumber \\
D(\mathcal{V}_{\lambda_2}, \mathbf{q}_{\lambda_2}) & \leq D(\mathcal{V}^*, \mathbf{q}^*) 
\label{app:inq_1}
\end{align}

where the second line is true because $B < R(\mathcal{V}_{\lambda_2}, \mathbf{q}_{\lambda_2})$ and $R(\mathcal{V}^*, \mathbf{q}^*) \leq B$. By the optimality of the solution $(\mathcal{V}^*, \mathbf{q}^*)$, we also know that:

\begin{equation}
D(\mathcal{V}^*, \mathbf{q}^*) \leq D(\mathcal{V}, \mathbf{q}), ~~ \;
\forall (\mathcal{V}, \mathbf{q}) \in \mathcal{S} 
\label{app:inq_2}
\end{equation}

where, $\mathcal{S}$ denotes the set of solutions that have a total rate lower than $B$. Note that $\mathcal{S}$ includes $(\mathcal{V}_{\lambda_1}, \mathbf{q}_{\lambda_1})$, since  $R(\mathcal{V}_{\lambda_1}, \mathbf{q}_{\lambda_1}) < B$. Combining the inequalities in (\ref{app:inq_1}) and (\ref{app:inq_2}), we can write:

\begin{align}
D(\mathcal{V}_{\lambda_2}, \mathbf{q}_{\lambda_2}) & \leq D(\mathcal{V}^*, \mathbf{q}^*) \leq D(\mathcal{V}_{\lambda_1}, \mathbf{q}_{\lambda_1}) \nonumber \\
\left| D(\mathcal{V}_{\lambda_1}, \mathbf{q}_{\lambda_1}) - D(\mathcal{V}^*, \mathbf{q}^*) \right| & \leq \left| D(\mathcal{V}_{\lambda_1}, \mathbf{q}_{\lambda_1}) -  D(\mathcal{V}_{\lambda_2}, \mathbf{q}_{\lambda_2}) \right|
\end{align}

which concludes the proof. $\square$

\bibliographystyle{IEEEtran}
\bibliography{ref}

\begin{thebibliography}{10}
\providecommand{\url}[1]{#1}
\csname url@samestyle\endcsname
\providecommand{\newblock}{\relax}
\providecommand{\bibinfo}[2]{#2}
\providecommand{\BIBentrySTDinterwordspacing}{\spaceskip=0pt\relax}
\providecommand{\BIBentryALTinterwordstretchfactor}{4}
\providecommand{\BIBentryALTinterwordspacing}{\spaceskip=\fontdimen2\font plus
\BIBentryALTinterwordstretchfactor\fontdimen3\font minus
  \fontdimen4\font\relax}
\providecommand{\BIBforeignlanguage}[2]{{%
\expandafter\ifx\csname l@#1\endcsname\relax
\typeout{** WARNING: IEEEtran.bst: No hyphenation pattern has been}%
\typeout{** loaded for the language `#1'. Using the pattern for}%
\typeout{** the default language instead.}%
\else
\language=\csname l@#1\endcsname
\fi
#2}}
\providecommand{\BIBdecl}{\relax}
\BIBdecl

\bibitem{tekalp95}
A.~M. Tekalp, \emph{Digital Video Processing}.\hskip 1em plus 0.5em minus
  0.4em\relax Upper Saddle River, NJ, USA: Prentice-Hall, Inc., 1995.

\bibitem{wang11}
P.~S.~P. Wang, \emph{Pattern Recognition, Machine Intelligence and
  Biometrics}.\hskip 1em plus 0.5em minus 0.4em\relax Springer Publishing
  Company, Incorporated, 2011.

\bibitem{kauff07}
P.~Kauff, N.~Atzpadin, C.~Fehn, M.~M\"{u}ller, O.~Schreer, A.~Smolic, and
  R.~Tanger, ``Depth map creation and image-based rendering for advanced 3dtv
  services providing interoperability and scalability,'' \emph{Image Commun.},
  vol.~22, no.~2, pp. 217--234, Feb. 2007.

\bibitem{shoham88}
Y.~Shoham and A.~Gersho, ``Efficient bit allocation for an arbitrary set of
  quantizers,'' in \emph{IEEE Transactions on Acoustics, Speech, and Signal
  Processing}, vol. 36, no.9, Sept. 1988, pp. 1445--1453.

\bibitem{kannan94}
K.~Ramchandran, A.~Ortega, and M.~Vetterli, ``Bit allocation for dependent
  quantization with applications to multiresolution and {MPEG} video coders,''
  in \emph{IEEE Transactions on Image Processing}, vol. 3, no.5, Sept. 1994.

\bibitem{liu05}
S.~Liu and C.-C.~J. Kuo, ``Joint temporal-spatial bit allocation for video
  coding with dependency,'' in \emph{IEEE Transactions on Circuits and Systems
  for Video Technology}, vol. 15, no.1, Jan. 2005, pp. 15--26.

\bibitem{kim05}
J.-H. Kim, J.~Garcia, and A.~Ortega, ``Dependent bit allocation in multiview
  video coding,'' in \emph{IEEE International Conference on Image Processing},
  Genoa, Italy, Sept. 2005.

\bibitem{cheung11tip2}
G.~Cheung, V.~Velisavljevic, and A.~Ortega, ``On dependent bit allocation for
  multiview image coding with depth-image-based rendering,'' in \emph{IEEE
  Transactions on Image Processing}, vol. 20, no.11, Nov. 2011, pp. 3179--3194.

\bibitem{wang06}
M.~Wang and M.~van~der Schaar, ``Operational rate-distortion modeling for
  wavelet video coders,'' \emph{IEEE Transactions on Signal Processing},
  vol.~54, no.~9, pp. 3505--3517, Sept 2006.

\bibitem{kaaniche14}
M.~Kaaniche, A.~Fraysse, B.~Pesquet-Popescu, and J.-C. Pesquet, ``Accurate
  rate-distortion approximation for sparse bernoulli-generalized gaussian
  models,'' in \emph{IEEE International Conference on Acoustics, Speech and
  Signal Processing}, May 2014, pp. 2020--2024.

\bibitem{bin12}
B.~Li, H.~Li, L.~Li, and J.~Zhang, ``Rate control by r-lambda model for {HEVC},
  {JCTVC-K0103},'' {Joint Collaborative Team on Video Coding} ({JCT-VC}),
  Changai, China, Oct. 2012.

\bibitem{woong13}
W.~Lim, H.~Jo, and D.~Sim, ``{JCT3V} – inter-view {MV}-based rate prediction
  for rate control of {3D} multi-view video coding, {JCT3V-F0166},'' {Joint
  Collaborative Team on Video Coding} ({JCT-VC}), Geneva, Switzerland, Oct.
  2013.

\bibitem{siwei12}
S.~Ma, J.~Si, and S.~Wang, ``A study on the rate distortion modeling for high
  efficiency video coding,'' in \emph{IEEE International Conference on Image
  Processing}, Sept 2012, pp. 181--184.

\bibitem{stuhlmuller00}
K.~Stuhlmuller, N.~Farber, M.~Link, and B.~Girod, ``Analysis of video
  transmission over lossy channels,'' \emph{IEEE Journal on Selected Areas in
  Communications}, vol.~18, no.~6, pp. 1012--1032, June 2000.

\bibitem{ortega98}
A.~Ortega and K.~Ramchandran, ``Rate-distortion methods for image and video
  compression,'' \emph{IEEE Signal Processing Magazine}, vol.~15, no.~6, pp.
  23--50, Nov 1998.

\bibitem{cheung07transcsvt}
G.~Cheung, W.-T. Tan, and C.~Chan, ``Reference frame optimization for
  multiple-path video streaming with complexity scaling,'' in \emph{IEEE
  Transactions on Circuits and Systems for Video Technology}, vol. 17, no.6,
  June 2007, pp. 649--662.

\bibitem{anavcip2014}
A.~De~Abreu, L.~Toni, T.~Maugey, N.~Thomos, P.~Frossard, and F.~Pereira,
  ``Multiview video representations for quality-scalable navigation,'' in
  \emph{IEEE Int. Conf. on Visual Communications and Image Processing},
  Valletta, Malta, Dec. 2014.

\bibitem{ana2015}
A.~D. Abreu, L.~Toni, N.~Thomos, T.~Maugey, F.~Pereira, and P.~Frossard,
  ``Optimal layered representation for adaptive interactive multiview video
  streaming,'' \emph{Journal of Visual Communication and Image Representation},
  vol.~33, pp. 255 -- 264, 2015.

\bibitem{Li2016}
S.~Li, C.~Zhu, Y.~Gao, Y.~Zhou, F.~Dufaux, and M.~Sun, ``Lagrangian multiplier
  adaptation for rate-distortion optimization with inter-frame dependency,''
  \emph{Circuits and Systems for Video Technology, IEEE Transactions on},
  vol.~26, no.~1, pp. 117--129, Jan 2016.

\bibitem{cheung00}
G.~Cheung and A.~Zakhor, ``Bit allocation for joint source/channel coding of
  scalable video,'' \emph{IEEE Trans. Image Processing}, vol.~9, no.~3, pp.
  340--356, Mar 2000.

\bibitem{sullivan98}
G.~Sullivan and T.~Wiegand, ``Rate-distortion optimization for video
  compression,'' in \emph{IEEE SP Magazine}, Nov. 1998.

\bibitem{cheung01}
G.~Cheung, ``Directed acyclic graph based mode optimization for {H.263} video
  encoding,'' in \emph{IEEE International Conference on Image Processing},
  Thessaloniki, Greece, October 2001.

\bibitem{garey99}
M.~R. Garey and D.~S. Johnson, \emph{Computers and Intractability: A Guide to
  the Theory of {NP}-Completeness}.\hskip 1em plus 0.5em minus 0.4em\relax
  Freeman, 1999.

\bibitem{Cormen2009}
T.~H. Cormen, C.~E. Leiserson, R.~L. Rivest, and C.~Stein, \emph{Introduction
  to Algorithms, Third Edition}, 3rd~ed.\hskip 1em plus 0.5em minus 0.4em\relax
  The MIT Press, 2009.

\bibitem{papadimitriou98}
C.~H. Papadimitriou and K.~Steiglitz, \emph{Combinatorial Optimization:
  Algorithms and Complexity}.\hskip 1em plus 0.5em minus 0.4em\relax Dover,
  1998.

\bibitem{Yue2009}
Y.-M. Chen, I.~Bajic, and P.~Saeedi, ``Coarse-to-fine moving region
  segmentation in compressed video,'' in \emph{Image Analysis for Multimedia
  Interactive Services, 2009. WIAMIS '09. 10th Workshop on}, May 2009, pp.
  45--48.

\bibitem{Yue2009_2}
Y.-M. Chen and I.~Bajic, ``Compressed-domain moving region segmentation with
  pixel precision using motion integration,'' in \emph{Communications,
  Computers and Signal Processing, 2009. PacRim 2009. IEEE Pacific Rim
  Conference on}, Aug 2009, pp. 442--447.

\bibitem{shark}
\BIBentryALTinterwordspacing
Shark multiview sequence. [Online]. Available:
  \url{http://www.fujii.nuee.nagoya-u.ac.jp/NICT/NICT.htm}
\BIBentrySTDinterwordspacing

\bibitem{Dancer}
D.~Rusanovskyy, P.~Aflaki, and M.~M. Hannuksela, ``Undo dancer 3{DV} sequence
  for purposes of 3{DV} standardization,'' in \emph{ISO/IEC JTC1/SC29/WG11
  MPEG2010/ M20028}, Geneva, Switzerland, March 2011.

\bibitem{goorts2014}
P.~Goorts, S.~Maesen, M.~Dumont, S.~Rogmans, and P.~Bekaert, ``Free viewpoint
  video for soccer using histogram-based validity maps in plane sweeping,'' in
  \emph{Proceedings of the Ninth International Conference on Computer Vision
  Theory and Applications (VISAPP 2014)}.\hskip 1em plus 0.5em minus
  0.4em\relax INSTICC, 2014.

\bibitem{karsten2014}
K.~Müller and A.~Vetro, ``Common test conditions of 3{DV} core experiments,
  {JCT3V-G1100},'' {Joint Collaborative Team on Video Coding} ({JCT-VC}), San
  José, US, January 2014.

\bibitem{sullivan12}
G.~J. Sullivan, J.~Ohm, W.-J. Han, and T.~Wiegand, ``Overview of the high
  efficiency video coding ({HEVC}) standard,'' in \emph{IEEE Transactions on
  Circuits and Systems for Video Technology}, vol. 22, no.12, Dec. 2012.

\bibitem{Liu2007}
Y.~Liu, Q.~Dai, Z.~You, and W.~Xu, ``Rate-prediction structure complexity
  analysis for multi-view video coding using hybrid genetic algorithms,'' in
  \emph{Proc. SPIE}, vol. 6508, 2007, pp. 650\,804--650\,804--8.

\bibitem{htm}
\BIBentryALTinterwordspacing
{HTM} 13.0 software. [Online]. Available:
  \url{\url{https://hevc.hhi.fraunhofer.de/svn/svn\_3DVCSoftware/tags/HTM-13.0/}}
\BIBentrySTDinterwordspacing

\bibitem{LarsLau12}
L.~Rakêt, L.~Roholm, A.~Bruhn, and J.~Weickert,
  ``\BIBforeignlanguage{English}{Motion compensated frame interpolation with a
  symmetric optical flow constraint},'' in
  \emph{\BIBforeignlanguage{English}{Advances in Visual Computing}}, ser.
  Lecture Notes in Computer Science.\hskip 1em plus 0.5em minus 0.4em\relax
  Springer Berlin Heidelberg, 2012, vol. 7431, pp. 447--457.

\bibitem{zhai05}
J.~Zhai, K.~Yu, J.~Li, and S.~Li, ``A low complexity motion compensated frame
  interpolation method,'' in \emph{IEEE International Symposium on Circuits and
  Systems}, Kobe, Japan, May 2005.

\bibitem{Schmeing2011}
M.~Schmeing and X.~Jiang, \emph{Depth Image Based Rendering}, P.~S. Wang,
  Ed.\hskip 1em plus 0.5em minus 0.4em\relax Springer Berlin Heidelberg, 2011.

\bibitem{Zinger2010}
S.~Zinger, L.~Do, and P.~H.~N. de~With, ``Free-viewpoint depth image based
  rendering,'' \emph{Journal Visual Communication and Image Representation},
  vol.~21, no. 5-6, pp. 533--541, July 2010.

\bibitem{hm}
\BIBentryALTinterwordspacing
{HM} 15.0 software. [Online]. Available:
  \url{\url{https://hevc.hhi.fraunhofer.de/svn/svn\_HEVCSoftware/branches/HM-15.0-dev/}}
\BIBentrySTDinterwordspacing

\bibitem{muller13}
K.~Muller, H.~Schwarz, D.~Marpe, C.~Bartnik, S.~Bosse, H.~Brust, T.~Hinz,
  H.~Lakshman, P.~Merkle, F.~Rhee, G.~Tech, M.~Winken, and T.~Wiegand, ``3d
  high-efficiency video coding for multi-view video and depth data,''
  \emph{Image Processing, IEEE Transactions on}, vol.~22, no.~9, pp.
  3366--3378, Sept 2013.

\end{thebibliography}

\end{document}